\documentclass[twocolumn]{emulateapj}

\usepackage{overpic}

\shorttitle{Multiple Populations in NGC 1851: Paper II}
\shortauthors{Cummings et al.}

\begin{document}

\title{Multiple Populations in NGC 1851: Abundance
Variations and UV Photometric Synthesis in the Washington
and HST/WFC3 Systems}


\author{Jeffrey D.\ Cummings\altaffilmark{1,2}, D.\ Geisler\altaffilmark{2}, and S.\ Villanova\altaffilmark{2}}

\altaffiltext{1}{Center for Astrophysical Sciences, Johns Hopkins University,
Baltimore, MD 21218, USA}
\altaffiltext{2}{Departamento de Astronom\'ia, Casilla 160, Universidad de Concepci\'on, Chile}


\begin{abstract}
The analysis of multiple populations (MPs) in globular clusters, both spectroscopically and photometrically, 
is key in understanding their formation and evolution.  The relatively narrow Johnson U, F336W, and Stromgren 
and Sloan u filters have been crucial in exhibiting these MPs photometrically, but in Paper I we showed that 
the broader Washington C filter can more efficiently detect MPs in the test case globular cluster NGC 1851.  
In Paper I we also detected a double MS that has not been detected in previous observations of NGC 1851.  We 
now match this photometry to NGC 1851's published RGB abundances and find the two RGB branches observed in C 
generally exhibit different abundance characteristics in a variety of elements (e.g., Ba, Na, and O) and in 
CN band strengths, but no single element can define the two RGB branches.  However, simultaneously considering 
[Ba/Fe] or CN strengths with either [Na/Fe], [O/Fe], or CN strengths can separate the two photometric RGB 
branches into two distinct abundance groups.  Matches of NGC 1851's published SGB and HB abundances to the 
Washington photometry shows consistent characterizations of the MPs, which can be defined as an O-rich/N-normal 
population and an O-poor/N-rich population.  Photometric synthesis for both the Washington C filter and the 
F336W filter finds that these abundance characteristics, with appropriate variations in He, can reproduce 
for both filters the photometric observations in both the RGB and the MS.  This photometric synthesis also 
confirms the throughput advantages that the C filter has in detecting MPs.
\end{abstract}

\section{Introduction:}

Globular clusters (GCs) have now in general been established, both photometrically and spectroscopically,
to have multiple populations (MPs) with differing compositions and possibly different ages.
Photometric detections of MPs in GCs
include Omega Cen (Bedin et~al.\ 2004), NGC 2808 (Piotto et~al.\ 2007), NGC 1851 (Milone et~al.\ 2008,
hereafter M08; Lee et~al.\ 2009a, hereafter L09; Han et~al.\ 2009, hereafter H09; Cummings et~al.\ 
2014, hereafter Paper I), M22 (Lee et al. 2009b), M4 (Marino et~al.\ 2008), and M2 (Lardo et~al.\ 2013; 
Milone et~al.\ 2015).  In the recent Piotto et~al.\ (2015) investigation, photometric detections of MPs are 
found in all 56 globular clusters observed with their special combination of filters (see below). 
In all of these clusters two or more red giant branches (RBGs), sub giant branches (SGBs), 
or even main sequences (MSs) have been observed.  

NGC 1851 is a noteworthy cluster because it photometrically has two RGBs 
(L09; H09) and SGBs (M08; H09) plus evidence for two MSs (Paper I).  Additionally, it has both a blue 
horizontal branch (BHB) and a red horizontal branch (RHB), where the Paper I analysis also suggested
the RHB has two sequences.  A further advantage to photometrically studying NGC 1851 is its very low 
reddening of E(B-V)=0.02 (Harris 1996), so there is no concern about the photometric effects of 
a large variable reddening.  
Spectroscopically, the two RGBs observed in NGC 1851 typically exhibit different abundances in a variety of 
elements, most strikingly in sodium (Na)
and barium (Ba) (Villanova et~al.\ 2010, hereafter V10; Carretta et~al.\ 2011a, hereafter Ca11) 
and in nitrogen (N) (Carretta et~al.\ 2014, hereafter Ca14).  Spectroscopic observations of 
molecular bands in NGC 1851 show its RGB has moderate variations in CH strength but large variations in 
CN strengths (e.g., Lim et~al.\ 2015), with no CN-CH anti-correlation.   There is also evidence for a 
quadrimodal CN distribution (Campbell et~al.\ 2012; Simpson et~al.\ 2017).  The two SGBs appear to correspond 
to high and low [Ba/Fe] (Gratton et~al.\ 2012a, hereafter G12).  Therefore, abundance 
differences likely play a key role in creating the separate sequences.   

Models to explain MPs of 
differing abundances include 1) An initial population formed and soon after ($<$1 Gyr) a second 
population formed from the gas contaminated by the ejecta of the high-mass stars of the first 
generation (see M08; Joo \& Lee 2013; Ventura et~al.\ 2009).  2) There was a merger of two GCs of 
slightly different age and composition (see Ca11).  The merger explanation is based on 
the Ca11 observation that between the two populations there is an apparent real spread in the 
heavy elements (e.g., iron (Fe)), not just in the light s-process elements.  This spread cannot simply be 
explained by two distinct episodes of star formation within one cluster.  3) Enriched material
was ejected from interacting massive binaries and rapidly rotating stars, and this material
accreted onto the circumstellar discs of young pre-main sequence stars formed at the same time
as these massive stars (Bastian et~al.\ 2013).  
Unlike the other models, this model only has a single star-formation burst that creates MPs 
based on variations in the amount of enriched material accreted, if any at all, on each individual 
cluster member.  This avoids the mass budget problem that plagues multiple star-formation burst models. 
However, a critical assessment of all current MP theories by Renzini et al. (2015) finds only the AGB 
pollution scenario to be possibly viable.

The key to photometrically distinguishing these MPs has been ultraviolet (UV) filters, where Sbordone 
et~al.\ (2011) and Carretta et~al.\ (2011b) have shown that realistic abundance differences in Carbon 
(C), Nitrogen (N), and Oxygen (O) greatly affect the UV filter bandpasses because of the strong CN, NH, 
and CH molecular bands present.  The ground-based detections of multiple sequences 
have used the Johnson U, Stromgren u, or sloan u filters, but these filters are narrow and inefficient and 
require a significant amount of large telescope time to observe most GCs well.  The Hubble Space Telescope (HST) 
has also been a powerful tool to photometrically analyze MPs with the F336W filter, which is comparable to 
Johnson U, and the F275W filter, which goes farther into the UV beyond what can be observed from the ground.  
Building on these filters, Milone et~al.\ (2013) and Piotto et~al.\ (2015) have defined the pseudo-colors 
C$_{\rm F275W,F336W,F410M}$ = (m$_{\rm F275W}$-m$_{\rm F336W}$)-(m$_{\rm F336W}$-m$_{\rm F410M}$) 
and the similar C$_{\rm F275W,F336W,F438W}$, respectively, terming these three the ''magic trio".  These pseudo-colors take advantage of the strong OH 
features in F275W's bandpass in combination with F336W's strong sensitivity to N, and this provides a key
method to photometrically distinguish most MPs.  Their team is carrying out a legacy survey of Galactic globulars 
in the magic trio and uncovering MPs in all of them, with each GC exhibiting unique behavior (e.g., Piotto et al. 
2015).  F275W, however, requires the use of HST, and this far into the UV it is even more difficult to acquire the 
appropriate signal on the typically cool GC stars, greatly increasing the need for very limited HST time.  
Furthermore, given the limited lifetime of HST and because no other current or planned space facility will 
be sensitive below the atmospheric cutoff, it is important to investigate other filters that can 
uncover MPs from the ground.

In Paper I we were thus motivated to consider other available tools without these limitations.  We demonstrated 
that the broader and more efficient Washington C filter was quite adept at photometrically detecting 
the NGC 1851 MPs with a ground-based 1-meter telescope and only moderate observation times.  
Color distribution analysis of the Washington photometry illustrated that these two populations 
are a dominant ($\sim$70\%) bluer population (in C-T1 and C-T2) that is narrow in color and a secondary 
($\sim$30\%) redder but partially overlapping population that is broadly distributed in color.
While the narrower band UV filters like Johnson U (H09) and Stromgren u (L09) appear to more 
cleanly separate the two populations in NGC 1851 into two distinct photometric branches, these two narrow
and farther UV filters required five and three times as much telescope time, respectively, to perform 
these observations (see Paper I for more details).

For this second paper in our analysis of NGC 1851, we look at the photometric variations in the Washington 
C filter and how they are connected to elemental abundance variations.  This is based on the detailed NGC 1851 
abundance analyses in Ca11, V10, Ca14, G12, Gratton et~al.\ (2012b, hereafter G12b), Yong et~al.\ (2015, 
hereafter Y15), and Lim et~al.\ (2015, hereafter L15). We also performed photometric synthesis and 
focused on the effects of variations in the three CNO abundances at constant C+N+O.  Additionally, we briefly 
looked at the photometric effects of variations in total C+N+O, metallicity, and in He.  We compare 
these effects in the C filter to the effects that identical variations have on the commonly used HST F336W 
filter, which is also very similar to Johnson U.  

In Section 2 we discuss the previous abundance results, including their trends and distributions.  In 
Section 3 we match previous abundances to out RGB photometry and analyze the abundance differences
between the two branches observed in Washington C.  In Section 4 we match previous
abundances to the SGB and turnoff stars.  In Section 5 we match previous abundances to HB stars.  
In Section 6 we synthesize for a representative RGB star the photometric effects that abundance 
variations have on C and F336W magnitudes.  In Section 7 we synthesize for a representative MS star 
the photometric effects that abundance variations have on C and F336W magnitudes.  
Lastly, in Section 8 we summarize our results and conclusions.

\section{Spectroscopic Abundances}

There have been multiple spectroscopic studies of NGC 1851 that have focused on its RGB stars (e.g., Ca11; V10; 
Campbell et~al.\ 2012; Ca14; Y15; L15), its turnoff and SGB stars (Pancino et~al.\ 2010; Lardo et~al.\ 
2012, hereafter L12; G12), and its HB stars (G12b).  All of these studies have shown that there is a broad 
spread of abundances (e.g., Na, O, Ba, Sr, Ni, and CN strengths) a possibly significant spread 
in Fe (Ca11), several abundance correlations and anti-correlations (e.g., O-Na and Ba-Sr), but
unlike most globular clusters it has no CN-CH anti-correlation (Pancino et~al.\ 2010; L12; L15; Simpson et~al.\ 2017).  
In our analysis we first looked at the abundance trends and distributions for each element and used these to 
define the high and low abundance ranges for each element.  
Matching these abundances to our Washington photometry allows us to analyze the abundance
variations in these stars in comparison to their placement on the photometrically observed MPs.  

\subsection{Abundance Trends}

In each element we have searched for abundance trends with T$_{\rm eff}$ to account for either true variations 
in surface abundances as stars undergo evolution or those caused by potential systematics in the analysis.  
Therefore, the distributions across a broad range of giants can more appropriately be analyzed.  Beginning 
with the analysis of 124 RGB stars from Ca11, 
we looked at the abundances of Fe, O, Na, Ca, Cr, and Ba.  Across the $>$1000 K range in T$_{\rm eff}$ 
there is no significant correlation with T$_{\rm eff}$ for Fe, Ca, and Ba.  For Na we find evidence for a 
minor anti-correlation of significance at greater than 99\% confidence.  Figure 1's Na panel shows this trend, which 
is minor relative to the overall observed scatter.  It is likely the result of systematics rather than a true 
abundance trend, but we have used this trend to divide the high and low Na abundances.

\begin{figure}[htp]
\begin{center}
\includegraphics[clip, scale=0.45]{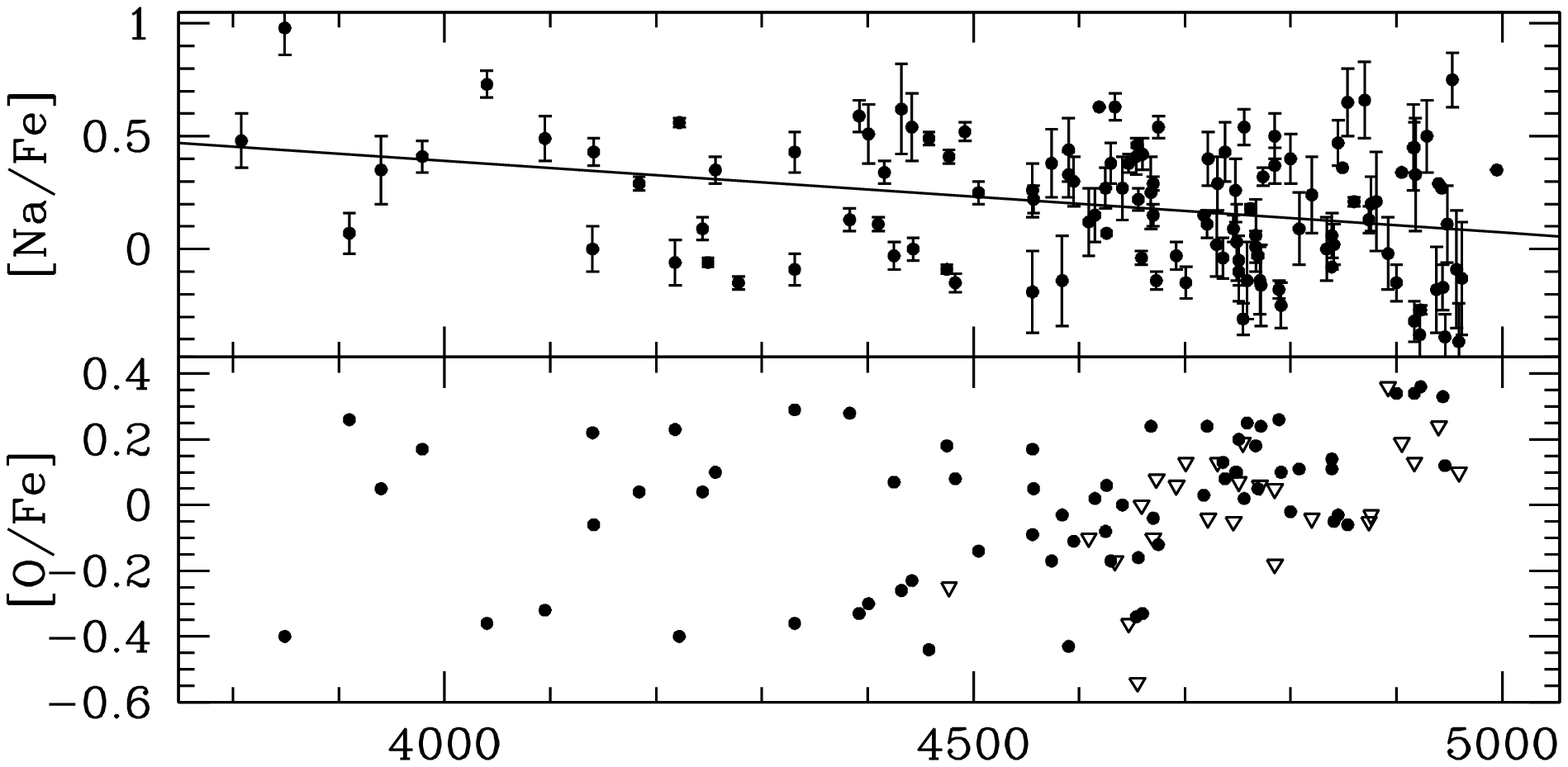}
\includegraphics[clip, scale=0.45]{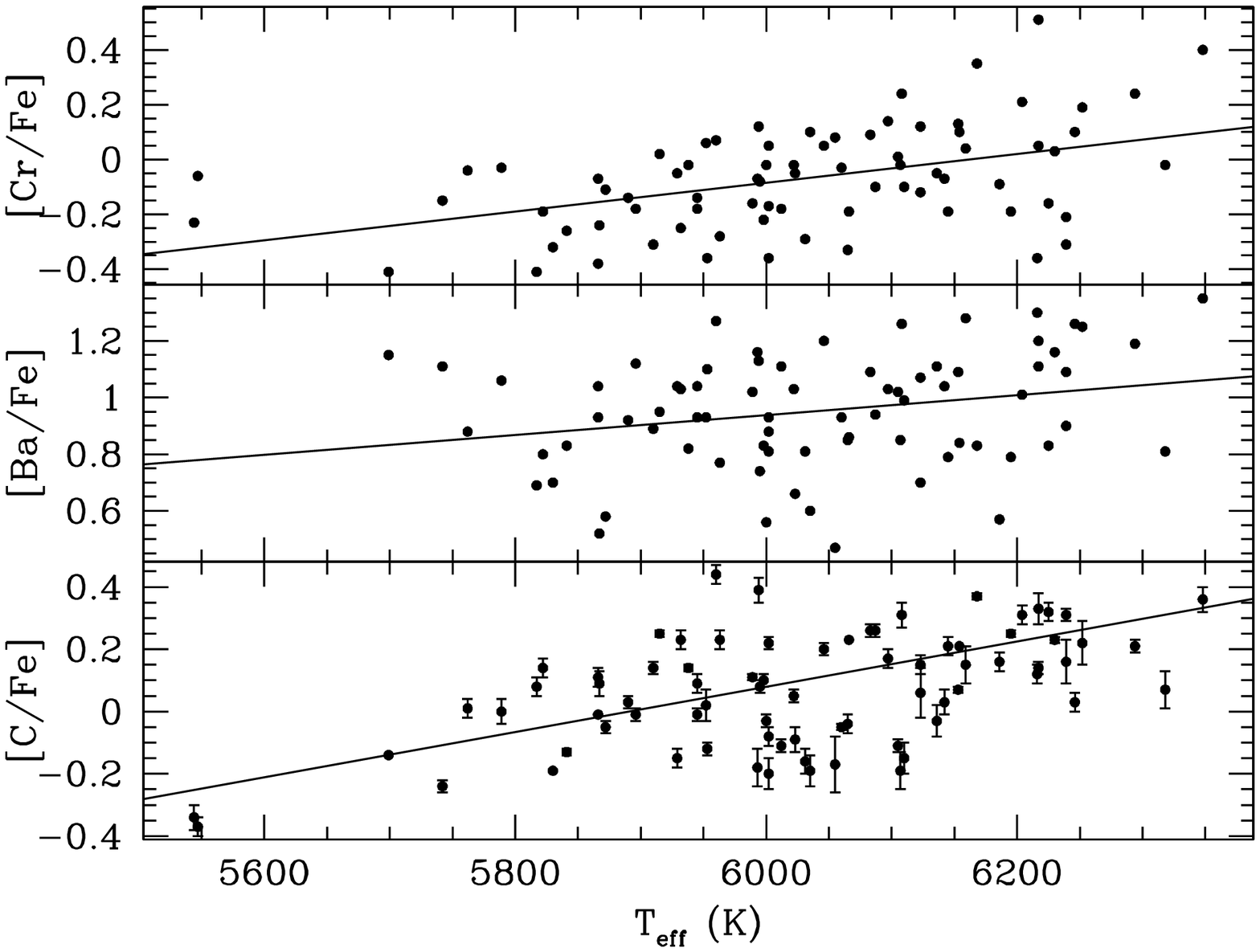}
\vspace{-0.75cm}
\end{center}
\caption{The upper two panels plot [Na/Fe] and [O/Fe] versus T$_{\rm eff}$ for the RGB stars observed in Ca11.  
Solid data points represent detections and open-inverted triangles represent O upper limits.  For Na the weak
but statistically significant correlation shown is Na$_{\rm cor}$ = 1.66 -- T$_{\rm eff}$ $\times$ 3.17$\times$ 
10$^{-4}$.  For O, in the cooler stars there is an apparent bimodal abundance distribution, but for the hotter 
stars the abundance trend at first appears to significantly increase.  However, this is because at these hotter 
T$_{\rm eff}$ the oxygen spectroscopic feature becomes very weak and unmeasurable in the O-poor stars.  This 
is illustrated by the large number of upper limits in the hotter stars.  Hence, there is no intrinsic 
trend with T$_{\rm eff}$ and O possibly has a bimodal distribution.  The lower three panels plot [Cr/Fe], 
[Ba/Fe], and [C/Fe] versus T$_{\rm eff}$ for the SGB stars observed in G12.  For Cr and Ba the minor but
statistically significant correlations shown are Cr$_{\rm cor}$ = --3.25 + T$_{\rm eff}$ $\times$ 5.27 $\times$ 10$^{-4}$
and Ba$_{\rm cor}$ = --1.17 + T$_{\rm eff}$ $\times$ 3.51 $\times$ 10$^{-4}$.  Lastly, the [C/Fe] correlation is
the most significant with C$_{\rm cor}$ = --4.29 + T$_{\rm eff}$ $\times$ 7.28 $\times$ 10$^{-4}$.  This may 
represent an intrinsic T$_{\rm eff}$ correlation caused by the increasing depth of
the surface convection zone along the SGB.  For these four abundance sets, these correlations define our rich and
poor populations for each element.} 
\end{figure}

The O abundances have an apparent trend where at cooler T$_{\rm eff}$ O appears to have a bimodal distribution 
while at higher T$_{\rm eff}$ there appears to be no O-poor stars.  Figure 1's O panel illustrates 
this, including upper limits for stars without detectable O shown as open-inverted triangles.  All of these 
upper limits are in the hotter RGB stars, and this is because the O spectral feature becomes increasingly 
weak in the fainter (hotter) RGB stars and was unmeasurable in the faint stars that are also 
O poor.  Therefore, this strongly suggests that the O abundances have no significant trend with 
T$_{\rm eff}$ and may have a bimodal distribution across the full T$_{\rm eff}$ range, but in the fainter 
(hotter) giants this bimodality is washed out by errors and the inability to measure the weakest [O/Fe].  

\begin{figure*}[htp]
\begin{center}
\includegraphics[scale=0.8]{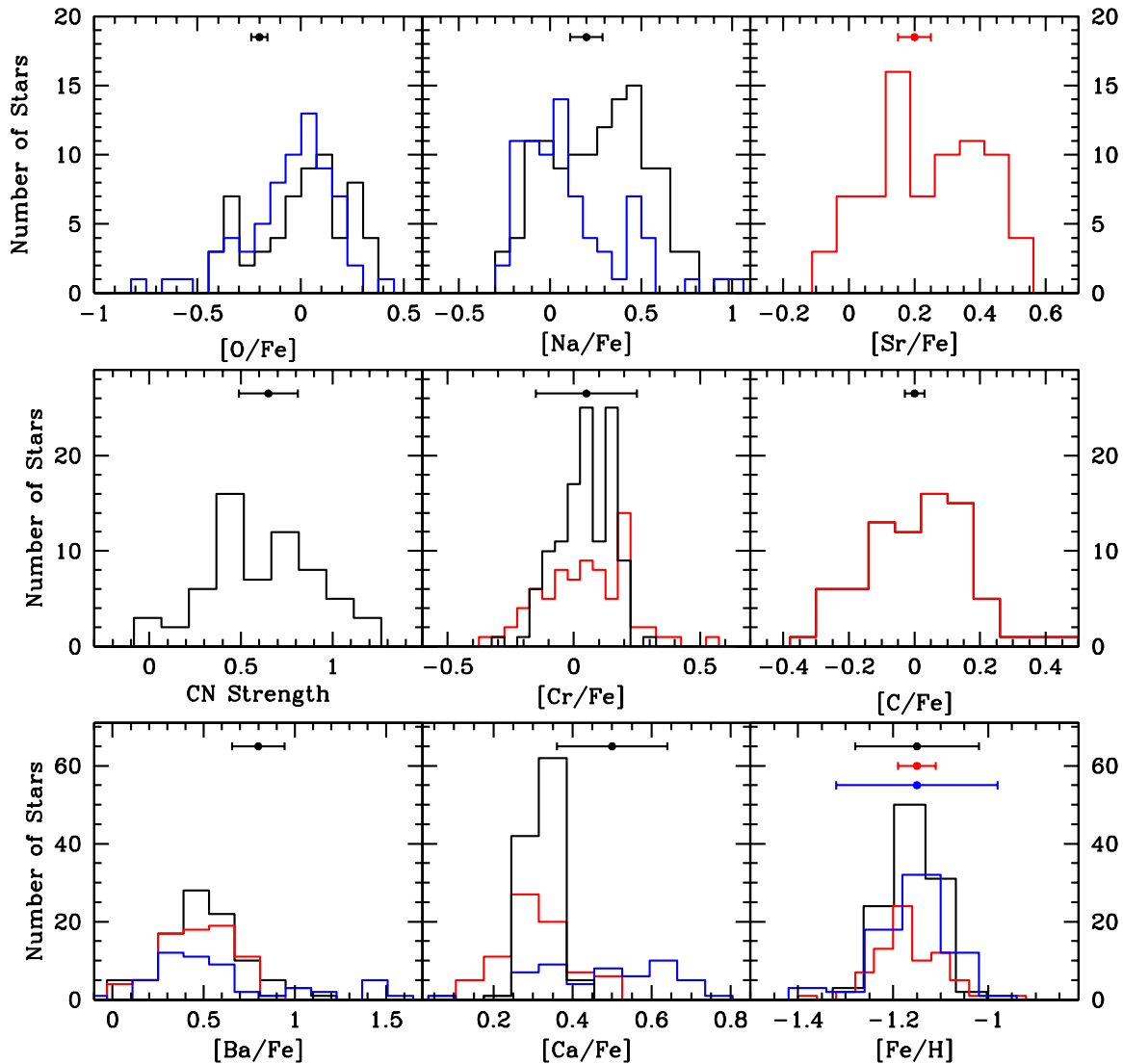}
\vspace{-0.43cm}
\end{center}
\caption{The abundance distributions for various elements are shown for the RGB (Ca11; black), the SGB (G12; 
red), and the HB/RGB (G12b; blue).  When available the typical abundance errors are shown above the distributions.
For display purposes, and to directly compare the characteristics of the abundance distributions, we have 
applied systematic offsets to place the distributions in line with those of the RGB when possible.  For [O/Fe] 
we applied -0.29 (G12b);  no [Na/Fe] offsets; for [Cr/Fe] we applied +0.16 (G12); for [Ba/Fe] we applied 
-0.45 (G12) and +0.3 (G12b); for [Ca/Fe] we applied -0.09 (G12) and +0.09 (G12b); no [Fe/H] offsets.}
\end{figure*}

Carretta et~al.\ (2014, hereafter Ca14) built further on the RGB analysis in Ca11 and looked at the CN 
strengths of 62 of the 124 RGB stars from Ca11.  In their analysis they state these CN strengths in terms of 
[N/Fe] with an assumed constant [C/Fe]=0.  While the RGB abundances from NGC 1851 in V10 
and Y15 suggest that [C/Fe] does not vary as significantly as [N/Fe] and [O/Fe] at a given magnitude, both find 
that [C/Fe] still does have important variations.  Additionally, surface abundance evolution due to 
deep mixing along the RGB will cause decreasing [C/Fe] in more evolved stars.  In the upper RGB, for example, 
[C/Fe] is more appropriately defined as $\sim$-0.8 dex rather than scaled solar.  Therefore, these [N/Fe] values 
from Ca14 will be used as a CN strength index, but we acknowledge that these [N/Fe] values 
likely trace true [N/Fe] variations.  Testing these CN strengths versus T$_{\rm eff}$ finds that there are no trends.

Building on these RGB abundances, we looked for similar correlations with T$_{\rm eff}$ in Fe, C, Ca, Cr, 
Sr, and Ba across the 77 SGB stars observed in G12 that span $\sim$800 K.  As for the cases in the RGB, we 
again did not find any trends for [Fe/H] and [Ca/Fe].  Sr was not analyzed in the RGB, but in the SGB we did 
not find any meaningful trend in [Sr/Fe].  In contrast to the RGB, there is evidence for weak but statistically 
significant trends in [Cr/Fe] and [Ba/Fe].  Both of these trends are shown in Figure 1.  Again, these are 
trends that are likely the result of minor systematics, and the inconsistency of the trends in the SGB and RGB 
supports this.  In any case, we have used these trends to help define our rich and poor abundances for these 
elements.  Lastly, the lower panel of Figure 1 shows [C/Fe] vs T$_{\rm eff}$ in the SGB from G12, which is the 
most scientifically interesting and strongest observed abundance-T$_{\rm eff}$ trend.  [C/Fe] steeply decreases 
when moving from the hotter (less evolved) stars up the SGB.  As G12 discuss, this trend is likely the result 
of mixing as the surface convection zones increase in depth along the SGB.

\subsection{Abundance Distributions}

Figure 2 shows the abundance distributions of nine elements or molecules of interest, where when available we 
have plotted the distributions derived from Ca11 (black; RGB stars), G12 (red; SGB stars), and G12b (blue; 
HB stars).  The median abundance errors (when published) are shown above the distributions.  For 
elements that we found 
correlations of significance in the previous section, we have adjusted these distributions to reflect that.  
Additionally, for elements that were analyzed in both Ca11 and G12, we have estimated and corrected any 
observed systematic differences in the distribution relative to the RGB abundances.  These applied 
corrections are given in the caption of Figure 2.  For the HB analysis, these systematics can be based 
directly on the overlapping samples of Ca11 and 
G12b, where G12b also looked at a number of RGB stars that included seven from Ca11.  The calculated 
systematics (Ca11 minus G12b) between these RGB stars found no meaningful difference in [Fe/H], 
a 0.09 dex difference in [Ca/Fe], and a 0.3 dex difference in [Ba/Fe], which have been applied in Figure 2.  
For [Na/Fe] and [O/Fe] the measured systematics are even larger at 0.31 and -0.56, respectively, but in 
Figure 2 these resulted in clear offsets between the observed distributions.  For display purposes we 
instead matched up the distributions with no offset in [Na/Fe] and a -0.29 offset in [O/Fe].  While the
[Na/Fe] and [O/Fe] differences between the RGB and HB may be real, for the purposes of this paper the cause 
of this systematic is not important.

\begin{figure*}[htp]
\begin{center}
\includegraphics[scale=0.75]{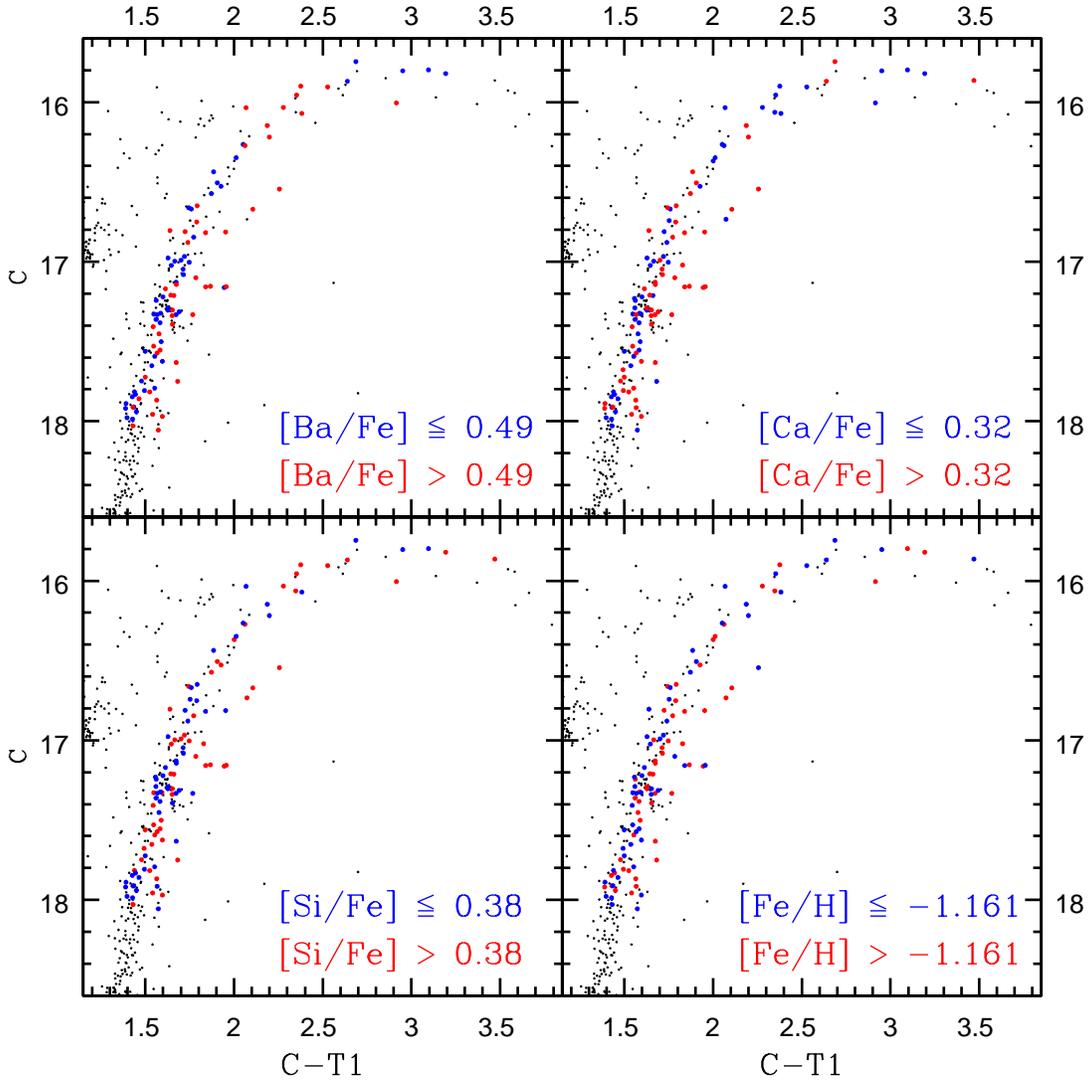}
\vspace{-0.43cm}
\end{center}
\caption{Matches of the Ca11 and V10 abundances to our Washington photometry for Ba, Ca, Si, and Fe.  In each
panel the analyzed stars rich in the element labeled are shown in red while the analyzed stars poor in 
that element are shown in blue.  The abundance ranges are defined in each panel.  
The photometric matches find that stars poor in Ba (and less so in Ca or Si) are consistently 
found to fall along the blue branch of the RGB.  In contrast, the stars that are rich in either Ba, Ca, or Si, 
are broadly distributed
in color and fall along both the blue branch as well as create the sparse red branch of the RGB.  The 
photometric difference between the Fe-rich and Fe-poor populations are not significant.}
\end{figure*}

Looking at the distributions themselves, while errors and number statistics limit the significance of some 
of these abundance variations, the distributions of many of the elements suggest a significant spread beyond 
that caused by errors.  Furthermore, some abundances even suggest a bimodal distribution (O, Na, Sr).  The 
significance of these bimodalities are further strengthened by being observed in more than one study.  For
example, the smaller sample of 15 RGB stars from V10, not shown here, also shows a clear bimodality in Sr.  
The distribution of [O/Fe] is of particular interest with (as seen in Figure 1) a small population of very 
O-poor RGB stars from Ca11, but here we also see that this population exists in the HB and RGB analysis of 
G12b, strengthening the significance of this bimodality.

\section{Matching Red Giant Branch Abundances to Photometry}

\subsection{Characterizing the Red Giant Branch Abundances}

Figure 3 matches the Ca11 abundances of Ba, Ca, Si, and Fe to our Paper I photometry of the two RGB 
populations in NGC 1851. We have also supplemented these abundances with three RGB star abundances from V10, 
which have been systematically adjusted to be consistent with the Ca11 abundances. This finds that the two 
photometric RGB branches have different abundance characteristics in Ba and Ca, but any differences
are less clear in Si and Fe.  The Ba-poor stars and the Ca-poor primarily 
fall tightly along the blue RGB, while the Ba-rich and Ca-rich stars are more widely 
distributed and cover both the blue and red RGB.  This photometric abundance distribution, most clearly
observed in Ba, is remarkably similar to the two photometric populations observed in Paper I.  Based 
solely on photometric color distribution analysis, we found that the C filter does not distinctly separate 
the MPs but creates a narrow blue population and an overlapping broader and redder second population.  

A more robust statistical analysis of these population distributions can be performed with a 
Kolmogorov-Smirnov test (KS-test).  In Figure 4 we look at the photometric C-T1 colors these 
stars relative to a mean RGB trend, and we present the cumulative distributions of the rich and poor populations 
for each element.  The KS-test looks at the measurement of the greatest separation (D) between each distribution, 
and based on the population sizes this provides the significance (p-value) of whether or not these different 
abundances correlate with distinct colors.  As is common we adopt p-values of $<$0.05 as 
representative of a statistically significant difference in color distributions between the two populations 
(i.e., that we can reject the null hypothesis that these abundance groups are derived from the same color 
distribution).

In Table 1 we give the KS-test statistics for each element and molecule analyzed (see Figures
3 to 6 for display of most of these elements and molecules).  This finds that in Figures 3 and 4 the 
distinct color distributions based on abundance in Ba and Ca are statistically significant.  While Si
shows a weaker distinction, the differences are still significant.  In contrast 
to this, the comparison of the Fe-rich and Fe-poor populations from Ca11 in the RGB gives a D of 0.2379 and 
p-value 0.078.  Therefore, while this does not rule out the significance of this possible [Fe/H] spread 
observed by Ca11, Fe does not correlate well with the different photometric branches.  Ca11 also found this 
when they matched the different [Fe/H] to the Stromgren photometry by Calamida et~al.\ (2007), where both the 
metal-rich and metal-poor stars showed similar double RGBs.  This is the foundation for their argument 
that NGC 1851 was formed from the merger of two different clusters.

\vspace{-0.4cm}
\begin{center}
\tablefontsize{\footnotesize}
\begin{deluxetable}{c c c c c}
\multicolumn{5}{c}%
{{\bfseries \tablename\ \thetable{} - Abundance Distribution KS-Test Statistics}} \\
\hline
Element & Poor Count & Rich Count & D & p-value\\
\hline
\hline
Ba & 51 & 47 & 0.4293 & 0.000\\
Ca & 53 & 56 & 0.3133 & 0.007\\
Si & 54 & 54 & 0.2593 & 0.043\\
Fe & 56 & 53 & 0.2379 & 0.078\\
Cr & 53 & 45 & 0.1379 & 0.711\\
Mg & 48 & 57 & 0.2007 & 0.218\\
O  & 36 & 36 & 0.3333 & 0.028\\
Na & 49 & 57 & 0.3863 & 0.000\\
CN & 47 & 42 & 0.5228 & 0.000\\
CH & 27 & 27 & 0.2222 & 0.466\\
\hline
\caption{For each element or molecule we matched to the RGB, we give the corresponding
number of rich and poor stars, (D) the greatest separation between the cumulative distributions of
the two populations, and the corresponding p-value where we adopt p$<$0.05 as significant.}
\end{deluxetable}
\end{center}

\begin{figure}[htp]
\begin{center}
\includegraphics[scale=0.44]{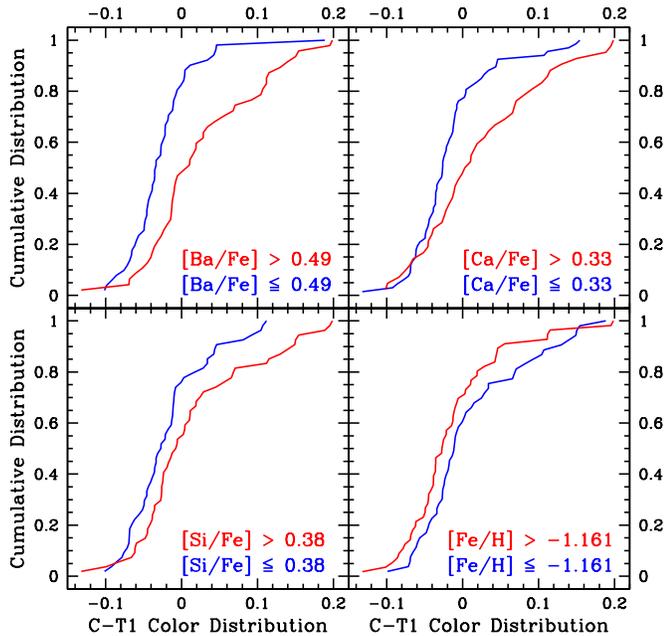}
\vspace{-0.43cm}
\end{center}
\caption{We analyze the statistical properties of the photometric abundance distributions in 
Figure 3.  Adopting the same abundance groups (and color schemes), we look at how the cumulative
distributions compare between the rich and poor populations for each element in C-T1 color space.
We derive relative C-T1 colors by fitting the mean RGB trend with the relation of 
C-T1=507.974-T1$\times$129.2119+T1$^2\times$12.44539-T1$^3\times$0.5351378+T1$^4\times$0.00864987.}
\end{figure}

In the upper panels of Figure 5 we match the key elements of O and Na (from Ca11 supplemented
with V10) to the RGB.
Like with Ba, Na clearly shows that nearly all Na-poor stars are consistent with the narrow
blue branch while the Na-rich population is broadly spread covering the blue branch but also creating
the red branch.  We similarly find a distinct distribution between the O-poor and O-rich stars, but 
for this element the blue branch is primarily O-rich (not poor like with Ba and Na) while the red
branch is predominantly created by O-poor stars.
Looking in more detail at [O/Fe], as we noted in Section 2.1, its abundance may have a bimodal 
distribution.  Much of this is washed out, however, by the difficult measurement of O in the faint and 
hotter RGB stars, which appears to have been unmeasurable in the O-poor stars in this regime.  Qualitatively 
consistent with this idea is that none of the faint red-branch RGB stars that were analyzed in Ca11 
have O detections in Figure 3; only the faint blue-branch RGB stars have O detections.  In the
upper panels of Figure 6 we look at the cumulative distributions of the O and Na abundance populations 
from Figure 5.

Why do these different abundance populations have such different photometric characteristics?
As described, the ``red RGB branch" is composed of second population stars but many second population 
stars also fall on the ``blue RGB branch".
On average the second population stars that fall on the red RGB branch are moderately
more Ba-rich, Na-rich, or O-poor than the second population stars that fall on the blue RGB branch.
However, several of the most Ba-rich, Na-rich, or O-poor stars still fall on the blue RGB branch.  This 
suggests it is not our adopted definitions of the abundance groups that create these overlapping distributions.
Ca11 similarly found photometrically overlapping populations in matches of their abundances to the Stromgren 
photometry of L09, and this suggests that MPs can create similar photometric characteristics in the C 
and Stromgren u filters.  

\begin{figure*}[htp]
\begin{center}
\includegraphics[scale=0.75]{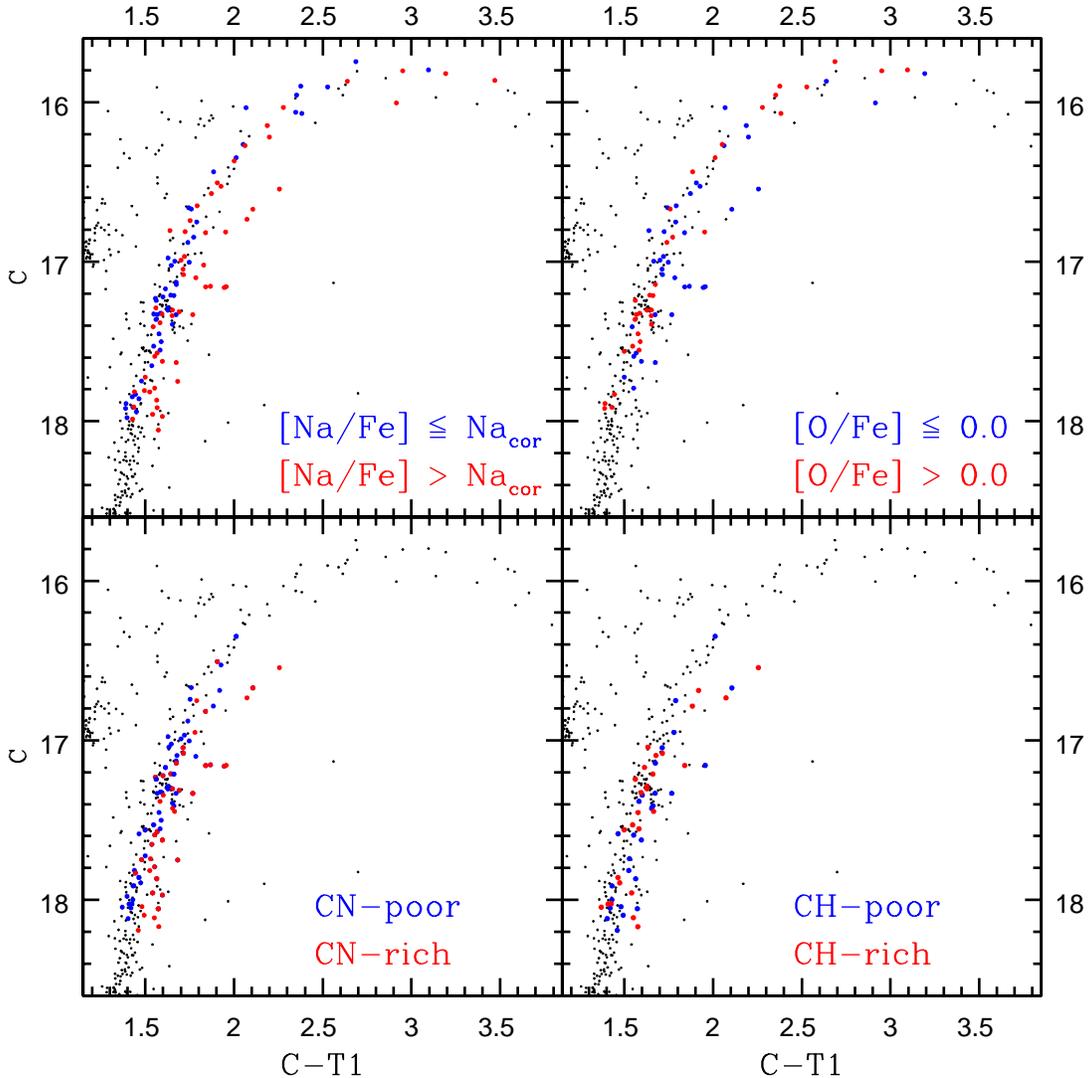}
\vspace{-0.43cm}
\end{center}
\caption{Matches of the Ca11 and V10 abundances to our Washington photometry for Na, and O, and matches 
of the CN and CH bands strengths from Ca14 and L15 to our Washington photometry.  In each
panel the analyzed stars rich in the element labeled are shown in red while the analyzed stars poor in 
that element are shown in blue.  The O abundance ranges are defined in the panel.  As seen 
in Figure 1, we use the [Na/Fe] correlation with T$_{\rm eff}$ to define the Na-rich and Na-poor stars.
For CN the CN-poor stars are [N/Fe]$\leq$-0.05 for the Ca14 abundances or $\delta$CN$\leq$0 for the L15 
abundances, and the CN-rich stars are [N/Fe]$>$-0.05 for the Ca14 abundances or $\delta$CN$>$0 for the 
L15 abundances.  For CH the CH-rich stars are $\delta$CH$>$0 and the CH-poor stars are $\delta$CH$<$0 from L15.
Like with Ba, the photometric matches find that stars poor in either Na or CN, or those that are rich 
in O, are consistently found to fall along the blue branch of the RGB.  In contrast, the stars that are 
rich in either Na or CN, or those that are poor in O, are broadly distributed
in color and fall along both the blue branch as well as create the sparse red branch of the RGB. The photometric
difference between the CH-rich and CH-poor stars are not significant.}
\end{figure*}

Carretta et~al.\ (2011b, hereafter Ca11b), Ca11, and more recently Ca14 have analyzed why the population 
that is typically poor in light s-process elements is photometrically very narrow and the population that is 
typically rich in light s-process elements is very broad in color and centered redward in the Stromgren 
filters.  Ca11b originally suggested a model based purely on variations in CNO abundances where 
the two stellar populations can be defined as C-normal and C-rich, which can recreate the general observed 
photometric characteristics.  However, in this paper we have focused instead on a photometrically similar 
model of two populations that are distinctly N-normal and N-rich in abundance.  This is based on four
recent abundance analyses: First, in Ca14 the redder RGB stars were found to be distinctly CN-rich 
and the bluer RGB stars were primarily CN-weak.  Supplementing this finding, the analysis by V10 suggested 
that the most clear abundance distinction between the giants in NGC 1851 are a Ba-rich and a Ba-poor 
population, but these Ba-rich and Ba-poor populations were also found to be N-rich/O-poor and 
N-normal/O-rich, respectively.  In V10 the two populations had a difference in [N/Fe] nearly as significant 
as [Ba/Fe].  Additionally, while the total V10 sample has moderate variations in [C/Fe], the Ba-rich and 
Ba-poor populations themselves show no significant difference in their mean [C/Fe].  The recent 
analysis by Y15 also looked at CNO in the two RGBs of NGC 1851 for 11 stars, and they find the same 
qualitative details as previous analyses.  Lastly, L15 looked directly at CN and CH band strengths 
in 62 RGB stars from NGC 1851.  While they found a moderate spread in CH band indexes, they found a significant 
spread in the observed CN band indexes.  Based on this, N (and O) appears to be a key element in
distinguishing the populations, which is expected due to the strong molecular bands sensitive to N in
the UV.  

\begin{figure}[htp]
\begin{center}
\includegraphics[scale=0.44]{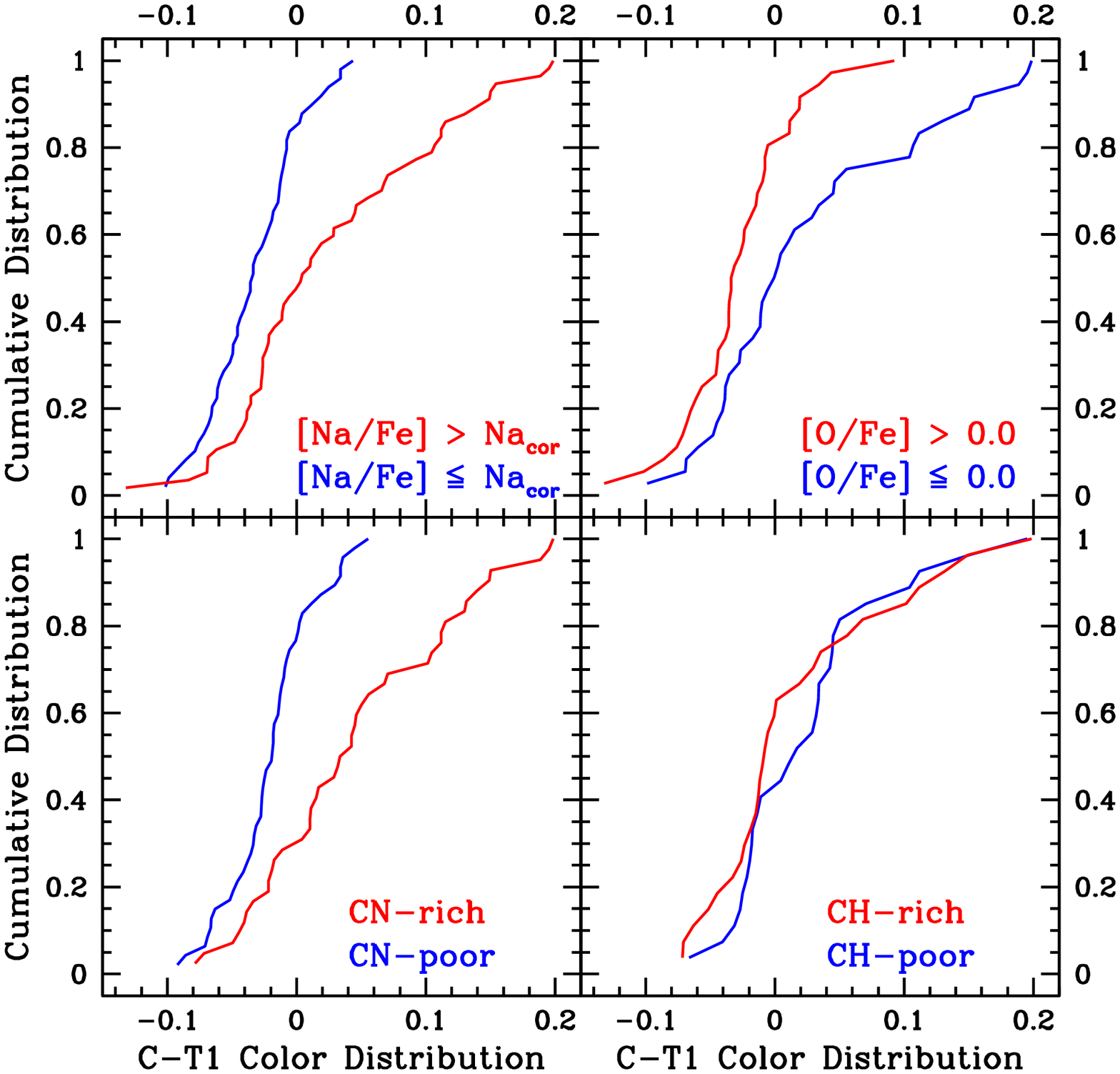}
\vspace{-0.43cm}
\end{center}
\caption{We analyze the statistical properties of the photometric abundance distributions in 
Figure 5.  Adopting the same abundance groups (and color schemes) we look at how the cumulative
distributions compare between the rich and poor populations for each element or molecule in C-T1 color space,
adopting the same mean RGB trend as in Figure 4.}
\end{figure}

Another key factor to consider between the abundances of the two RGB branches is whether the observed 
variations in C, N, and O still result in a constant total C+N+O abundance.  The analysis in V10 
found that both populations did have a consistent log $\epsilon$(CNO)$\sim$8.00\textsuperscript{3}.  
Comparison of V10 to the results of Y15 finds encouraging consistency in the blue RGB, 
but the Y15 analysis of the red RGB stars finds they are nearly a factor of 10 ($\sim$0.9 dex) richer 
in N than the blue RGB.  V10 only found them twice as rich in N ($\sim$0.35 dex) with respect to the blue 
RGB.  A comparison of the CN strengths in Ca14 similarly suggests the CN-rich stars are $\sim$0.4 dex 
richer in N.  This far more significant increase in N found
by Y15 also results in a distinct log $\epsilon$(CNO) abundance between the two branches,
with the red RGB branch being $\sim$0.5 dex richer.   In our synthetic magnitude analysis (see Sections 6 
and 7) we have adopted a constant log $\epsilon$(CNO) of 8.0, but consistent with Y15 we also briefly 
consider the effects of a significant increase in [N/Fe] in the red branch that results in a distinct 
log $\epsilon$(CNO) of 8.5 for the red.  

To look more directly at the effects of CNO and the corresponding molecular bands, we first use 
the CN strengths from Ca14 (represented by [N/Fe]) and L15 (represented by $\delta$CN).  These CN band 
strengths are useful because they (with the CH and NH bands) are the primary cause of the observed
photometric differences in these stars.  The lower-left panel of Figure 5 shows the CN abundances matched 
to our photometry with blue representing CN-poor ([N/Fe]$\leq$-0.05 for the Ca14 abundances or 
$\delta$CN$\leq$0 for the L15 abundances) and red representing CN-rich ([N/Fe]$>$-0.05 for the Ca14 abundances or 
$\delta$CN$>$0 for the L15 abundances).
Consistent with what we saw with Ba and Na abundances, we similarly see that the CN-poor stars are narrowly
distributed and are the primary component of the blue-RGB branch, while the CN-rich stars are
broadly distributed and create the red-RGB branch but also are heavily overlapping with the blue RGB.
In the lower-left panel of Figure 6 we show the cumulative
population distributions for these CN-rich versus CN-poor stars, and the KS test finds CN gives a 
D of 0.5228 with a p-value of 0.000.  This is the largest D given for a single abundance, but the two
populations are still heavily overlapping.
This suggests that while CN bands play a critical role, other factors, like the CH and NH bands, 
also must play a role in NGC 1851 and the C filter.  

The CH indices from L15 do not show the broad variations observed in CN indices, but they still have
meaningful variation.  In the lower-right panel of Figure 5 we have matched these CH band strengths to
our RGB photometry with CH-rich ($\delta$CH$>$0) in red and CH-poor ($\delta$CH$<$0) in blue.  This finds 
that there is no clear matching of either CH population with the two RGB branches.  In the lower-right panel 
of Figure 6 the cumulative distributions of the CH-poor and CH-rich populations are not distinct.  This is 
consistent with the lack of anti-correlation between CH and CN observed in Pancino et~al.\ (2010), L12, and L15, 
and with the two RGB populations observed in V10 showing no significant difference in [C/Fe].   
Therefore, there is a moderate spread in [C/Fe] throughout NGC 1851, but the two populations are not meaningly 
different in their [C/Fe] abundance distributions.

\footnotetext[3]{We adopt standard abundance notation where for a given element X,
log $\epsilon$ = log (N$_{\rm X}$/N$_{\rm H}$) and [X] = log $\epsilon$(X)$_{\rm star}$ - 
log $\epsilon$(X)$_\odot$.}

\subsection{Characterizing the Two Photometric Red Giant Branches}

What could cause the large color range observed only in the second RGB population?  A possible explanation is 
that NGC 1851 has two populations with distinct [O/Fe], and when adopting no variation in C+N+O, at a 
constant [O/Fe] the variations in [C/Fe] will be anti-correlated with [N/Fe]. 
Focusing on CN molecular bands, which dominate in the C filter bandpass, variations in [C/Fe] do 
not greatly affect the CN strengths of the O-rich (blue RGB) stars because the strengths are limited 
by their weaker N abundance.  Conversely, for the O-poor (red RGB) stars these CN bands are more significantly 
affected by [C/Fe] variations because they are N-rich and the N abundance is no longer a limiting factor.
The first population stars 
are O-rich (and typically N-normal) and will cover a tight color range while the second population stars are
O-poor (and typically N-rich) and will be more broadly distributed in color.  Therefore, the sparser and distinct
red RGB branch (which is only the reddest part of the second population) is composed of only the N-rich, C-rich, 
and O-poor stars, which have the strongest CN bands but also the strongest CH and NH bands.  In contrast, the 
blue RGB is composed of all the O-rich (first population) stars in addition to the 
O-poor stars that are also C-poor, which all have relatively weaker CN bands and either weak CH or NH 
bands, if not both. 

\begin{figure*}[htp]
\begin{center}
\includegraphics[scale=0.75]{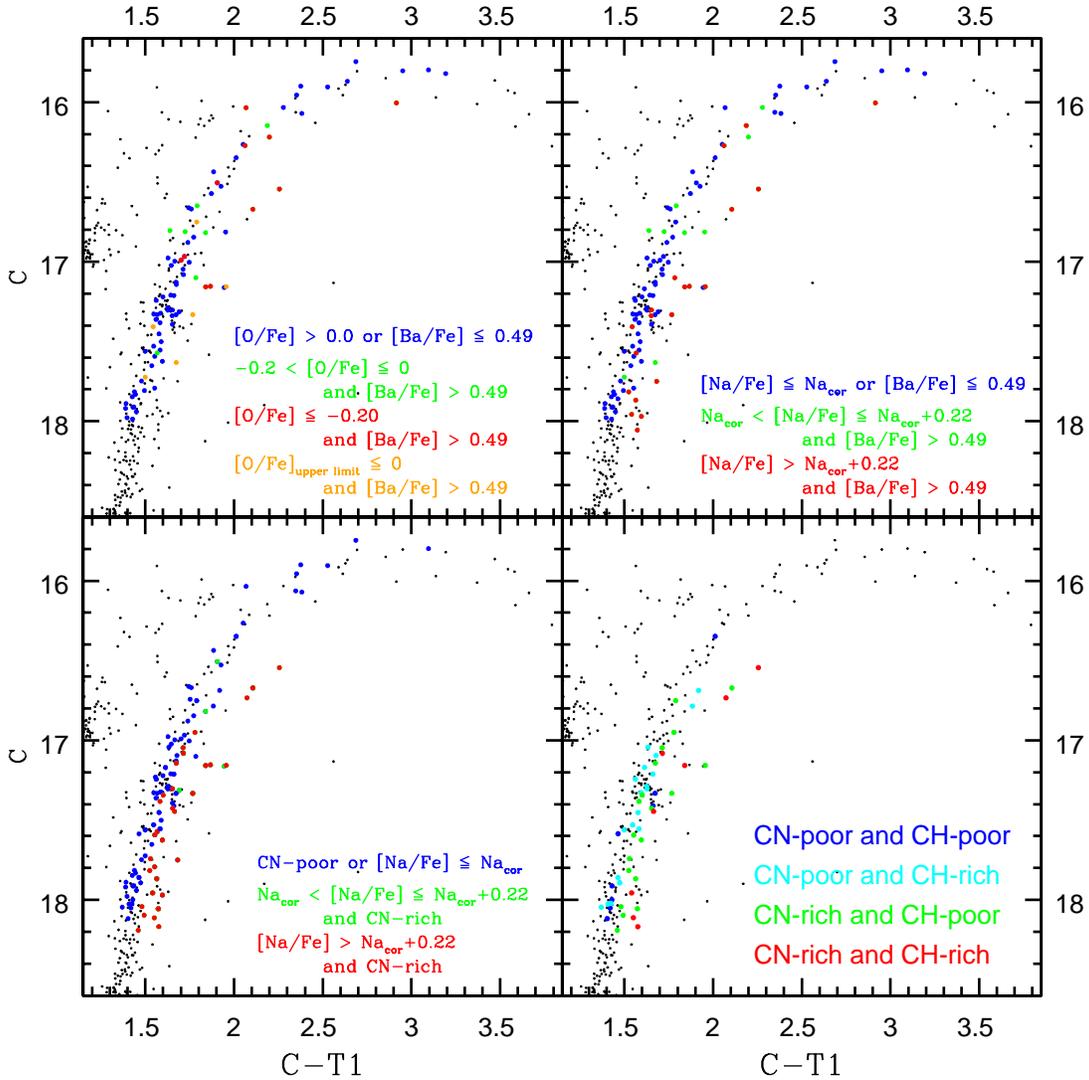}
\vspace{-0.43cm}
\end{center}
\caption{In these four panels we look at various methods to distinctly differentiate the abundances of the
red and blue RGB branches in the Washington filters.  In the upper-left panel we see that considering O and 
Ba simultaneously has some success at differentiating the abundance characteristics of the two branches.  This shows
that the blue RGB are stars that are either O-rich or Ba-poor (blue), and the red
RGB are stars that are both very O-poor and Ba-rich (red).  In the upper-right
panel we consider the more complete abundances of Na with Ba.  This more clearly differentiates
the blue RGB as stars that are either Na-poor or Ba-poor (blue), and the red RGB as
stars that are both very Na-rich and Ba-poor.  In the lower-left panel we consider CN-poor or Na-poor stars (blue),
CN-rich and Na-rich (green), and CN-rich and very Na-rich (red).   This most distinctly separates the two
photometric branches.  Lastly, in the lower-right panel we consider CN together with CH, where CH by itself did not
show any meaningful photometric differences.  Consistent with this, for CN-poor stars there is no significant difference
between CH-rich and CH-poor stars.  However, this comparison finds that the CN-rich stars that are consistent with
the blue branch are primarily CH-poor and the reddest giants are both CN-rich and CH-rich.}
\end{figure*}

\begin{figure}[htp]
\begin{center}
\includegraphics[scale=0.44]{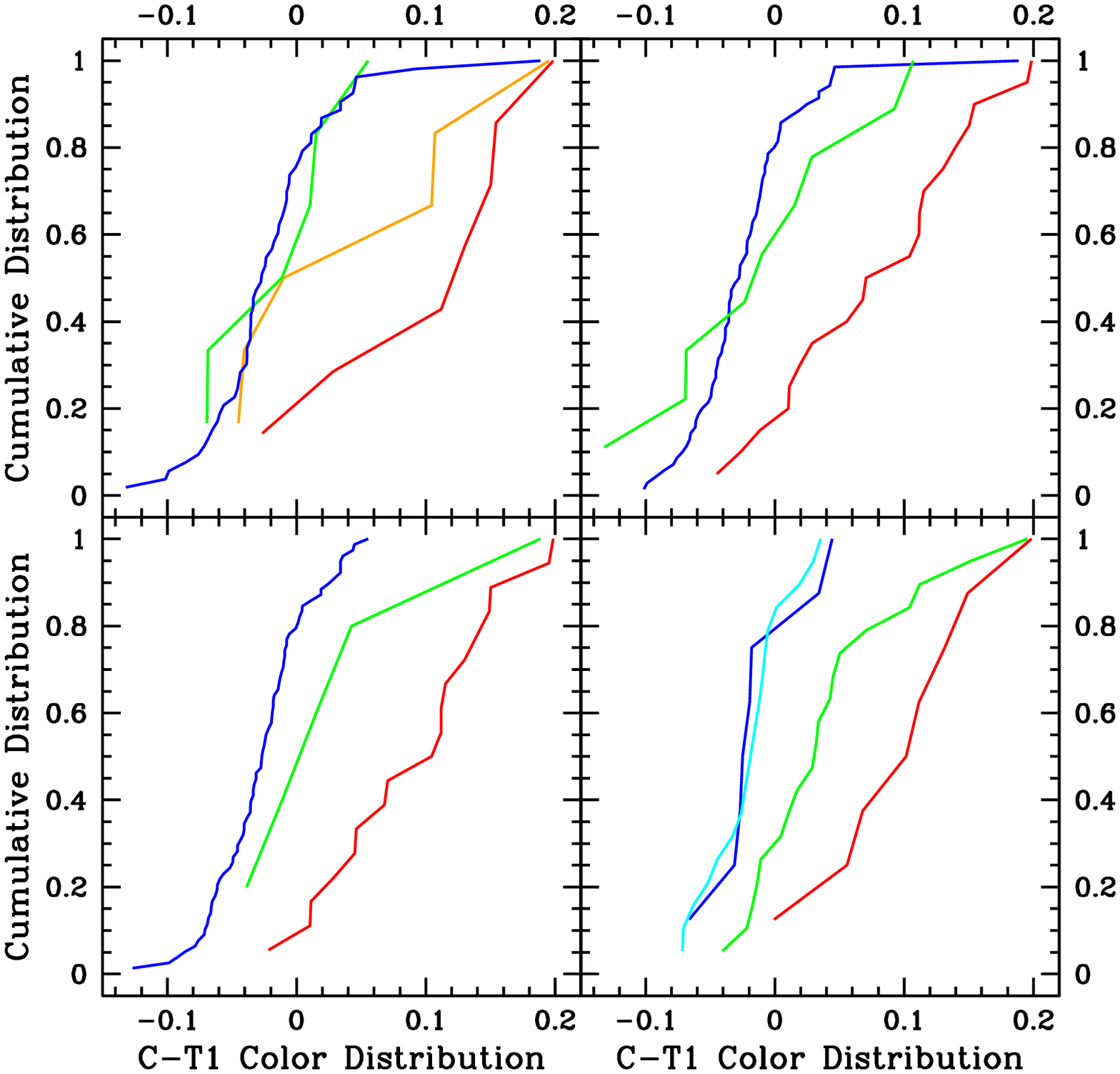}
\vspace{-0.43cm}
\end{center}
\caption{We analyze the statistical properties of the photometric abundance distributions in 
Figure 7.  Adopting the same abundance groups (and color schemes) we look at how the cumulative
distributions compare between the defined abundance populations in C-T1 color space,
adopting the same mean RGB trend as in Figure 4.}
\end{figure}

Another key element for differentiating the two populations is Ba.  The Ba absorption itself 
is not significant enough to affect the observed flux, but its differentiating power may be related to
a connection between Ba and N, where V10 found that the Ba-rich stars were also more N-rich.  
This is also found by our matching of the [Ba/Fe] from the much larger RGB sample in Ca11 to their CN 
abundances from Ca14; the Ba-rich stars are typically found to be more CN-rich by $\sim$0.15 dex.  
However, there is significant dispersion in the comparison and only evidence for a weak but statistically 
significant ($>$95\% confidence) [Ba/Fe] and CN correlation.  
Whatever connection Ba may have, looking at the upper-left panel in Figure 3 shows that the Ba-poor 
stars show little scatter and almost all of them fall directly on the blue RGB, while the Ba-rich stars 
stars show a very large scatter covering both the red and the blue RGB.  

\begin{figure*}[htp]
\begin{center}
\includegraphics[scale=0.85]{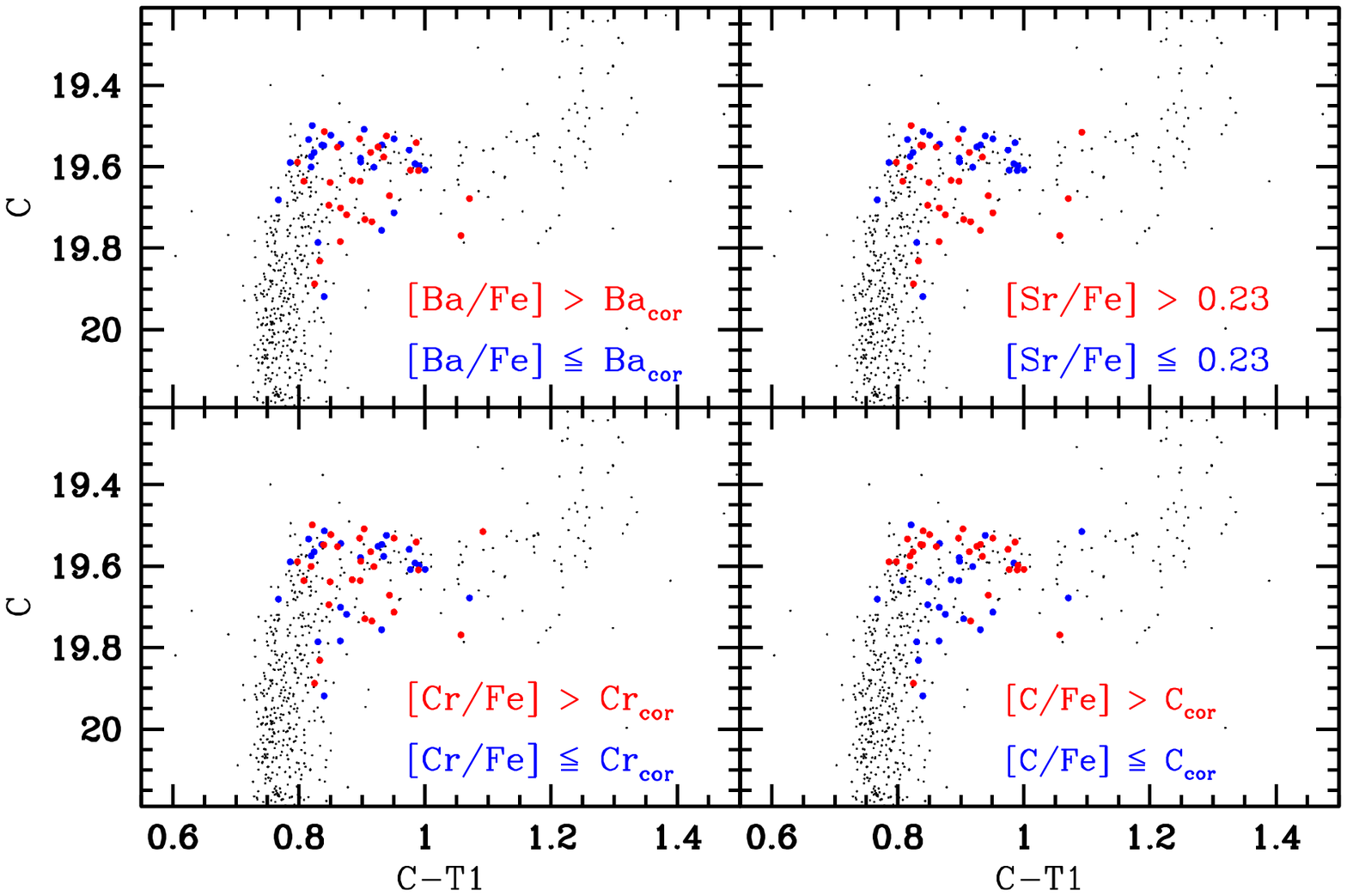}
\vspace{-0.43cm}
\end{center}
\caption{Similar to Figures 3, 5, and 7, we now match SGB abundances from G12 to our Washington photometry
of NGC 1851.  In the upper-left panel we see that again, like in the RGB, Ba-poor (blue) stars nearly all
fall tightly along the brighter (bluer) SGB while the Ba-rich (red) stars are more broadly distributed
in color and create the sparser fainter (redder) SGB but still also fall on the bright SGB.  Sr also creates a 
similar distribution.  Cr, however, does not create a clear distinction between the two SGB branches.  In
the lower-right panel we look at the C abundances and show that the C-rich stars predominantly fall
along the bright SGB while the C-poor stars are more broadly distributed.  (See Figure 1 for the definitions
of Ba$_{\rm cor}$, Cr$_{\rm cor}$, and C$_{\rm cor}$.)}
\end{figure*}

In Figure 7 we look at four approaches to distinctly characterizing the abundances of the red and 
blue RGBs.  The upper-left panel simultaneously looks at both [O/Fe] and [Ba/Fe].  We have colored all of the 
Ba-poor (N-normal) stars blue and all of the O-rich stars blue because as discussed these stars should 
be CN-poor and either CH-poor or NH-poor.  We clarify that these blue stars are either Ba-poor or O-rich
and not necessarily both, and we include Ba-poor stars that do not have
an O abundance and vice-versa.  We have used three colors to mark the stars that are both
Ba-rich and O-poor: green represents the Ba-rich and moderately O-poor stars, red represents the 
Ba-rich and very O-poor stars, and orange represents the Ba-rich stars with O upper limits that define 
them to be at least moderately O-poor if not very O-poor.  
Quite remarkably, the blue data are in agreement with the well defined blue RGB, while the green data show 
a moderately redder distribution typically falling on the red edge of the blue RGB and extending slightly 
redder, and the red data are mostly consistent with the red RGB.  As should be expected the orange data 
are consistent with both the green and red data.  This is promising but the 
matches are not perfect because two of the stars that clearly belong to the red RGB are colored blue.  We 
also note the one colored red data point that lies well above the RGB.  Its characteristics strongly suggest 
that it is an asymptotic giant branch star.  This, rather than its abundances, explains its photometric 
deviation above the RGB.  In the upper-left panel of Figure 8 we show the 
cumulative population distributions of these four abundance groups.  

The one limitation of the Ca11 O abundances is that O is a challenging measurement.  This resulted in 52 of 
the 111 stars we have matched to our photometry having only an upper limit or no O abundance information at 
all.  Therefore, because Na has a well established anti-correlation with O and Ca11 has 108 measured Na 
abundances, we have analyzed a similar combination of Na and Ba abundances.  This is shown in the upper-right
panel of Figure 7 with all of the Ba-poor or Na-poor stars colored blue.  The Ba-rich
and Na-rich stars are grouped into Ba-rich and moderately Na-rich as green or Ba-rich and very Na-rich as 
red.  This provides the most striking plot we have seen so far, where the blue data are consistently in 
agreement with the well defined blue RGB, while the green data show a moderately redder distribution 
typically falling where the red and blue RGBs meet, and the red data are consistent with the red RGB.  
Similarly, in the upper-right panel of Figure 8 we show the cumulative population distributions 
of these three abundance groups.
Unlike with O, use of Na provides both a larger sample and abundance information extending to the 
faintest observed red RGB stars, where we still see abundance patterns consistent with the bright RGB 
stars.  Overall, we have greatly increased the number of stars and find a result consistent with the proposed
model.

As with Ba, the O (Na) abundances in combination 
with CN may tell us more.  Again, the O abundances are limited in number, but the Na abundances will reliably 
be indicative of O.  The lower-left panel of Figure 7 shows the combination of the CN strengths 
and Na abundances, where we color stars that are both CN-rich and very Na-rich (very O-poor) red, the 
CN-rich and moderately Na-rich stars green, and all stars that are either CN-poor stars or Na-poor blue.  
We now see that when considering the O abundances (represented by Na) in addition to the CN strengths,
the red data and the blue data agree very well with the red RGB and the blue RGB, respectively, and
the green data primarily falls where the red and the blue RGB meet.  This is displayed more
clearly in the lower-left panel of Figure 8, where there is almost no color overlap between the red and
blue population distributions.  This result implies that because of 
the possible lack of large C+N+O variations, the CN-rich and Na-poor (O-rich) stars will not be both very C-rich 
and N-rich, so their moderately strong CN strengths will be balanced out by either weak CH or NH lines.  
Additionally, the CN-poor and Na-rich stars are O-poor but likely have band strengths limited by weak
N or weak C.  Only the CN-rich and the Na-rich (O-poor) stars will be significantly rich enough in both C 
and N to also be CH-rich and NH-rich, leading only these stars to be significantly fainter in the C magnitude.

Lastly, in the lower-right panels of Figures 7 and 8 we consider CH and
CN band strengths together.  Here stars that are both CN-poor and CH-poor are blue, those that
are both CN-poor but CH-rich are cyan, those that are both CN-rich and CH-poor are magenta, and lastly those
that are both CN-rich and CH-rich are red.  Remarkably consistent with our explanation given at the beginning
of this section, and most clearly shown in the lower-right panel of Figure 8, for the CN-poor stars the variations in CH strengths 
(i.e., [C/Fe]) have no meaningful affect on the resulting color.  This is not the case for the CN-rich stars, 
where all CN-rich stars consistent with the blue RGB are also CH-poor, while the red RGB is composed of nearly 
all of the stars that are both CN-rich and CH-rich.

\section{Matching Turnoff/Subgiant Branch Abundances to Photometry}

We have also matched the SGB and turnoff abundances of G12 to our photometry.  While the lack of [O/Fe] 
and [Na/Fe] measurements for these stars limit us from performing the more detailed analysis we did with the 
RGB stars, we still detect that in several elements the two branches exhibit different abundance characteristics.  
In Figure 9 both [Ba/Fe] and [Sr/Fe] show similar patterns as those observed for Ba in the RGB.  The Sr-poor 
stars and Ba-poor stars are both concentrated on the brighter and blue branch while the Sr-rich stars and the 
Ba-rich stars are more broadly distributed and create the sparse fainter and red branch but also compose part of the bright 
branch.  Sr and Ba are the two elements that both G12 (in the SGB) and V10 (in the RGB) show to be strongly 
correlated.  The Cr-rich and Cr-poor stars do not show a strong photometric distinction; this illustrates the 
importance of adopting the observed trends (see Figure 1) to define rich and poor populations because adopting
a constant separation for Cr-rich and Cr-poor does result in an apparent photometric distinction in the SGB.  
These Ba and Sr distributions suggest that in the C filter, as on the RGB, these are not two 
photometrically distinct SGB sequences representing two cleanly separated populations, but that photometrically 
the two populations are a narrow and a broad population that overlap each other.  
The broad second population extends significantly farther to the red and creates a distinctly redder branch, but again 
this clearly redder group is not the entire second population.  

\begin{figure*}[htp]
\begin{center}
\includegraphics[scale=0.75]{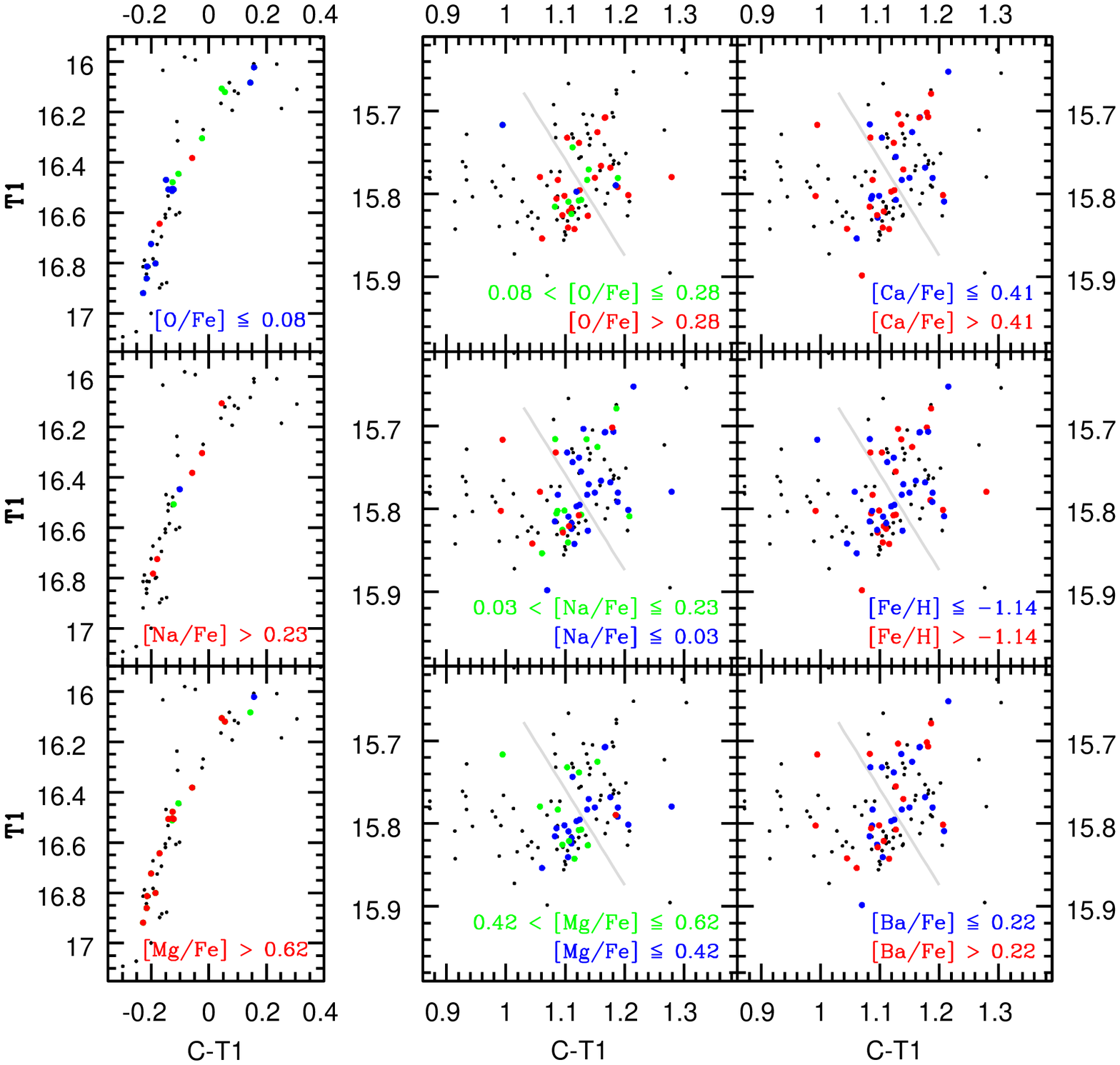}
\vspace{-0.43cm}
\end{center}
\caption{Matches of the HB abundances from G12b to our Washington photometry
of NGC 1851.  The left grid of three panels shows the BHB and the right grid of 6 panels shows the 
RHB, with its two potential sequences: a faint (bluer) sequence and a bright (redder) sequence, which we
have divided by a solid grey line.  O, Na,
and Mg have both BHB and RHB abundances and their abundances distributions are very broad.  Therefore, 
we have grouped them in three abundance groups with blue being poor, green being intermediate, and 
red being rich.  This clearly illustrates the abundance distinctions between the RHB and BHB.  Looking
closely at the RHB itself finds that there are no clear abundance distinctions between the two 
apparent RHB sequences in any of these six elements.}
\end{figure*}

G12 also have direct measurement of the C abundances from the CH bands for the SGB stars, which we
discussed a temperature trend for in Section 2 and Figure 1.  Using the correlation from Figure 1 to define our
C-rich and C-poor populations and matching these abundances to our photometry shows that the faint 
SGB is predominantly C-poor while the bright SGB is predominantly C-rich.  This is in contrast to what 
we would infer based on the RGB abundances and CH indices, but the bright SGB is on average only 
$\sim$0.1 dex richer in C.  This difference is relatively minor compared to the total range of
C abundances, when correcting for the trend with T$_{\rm eff}$, spanning $\sim$0.6 dex.  To expand
on this further, we can define the two SGB populations based on their [Ba/Fe], and we find that the Ba-rich (second
population) and Ba-poor (first population) stars have on average no meaningful difference in C.

We also acknowledge the work in L12, which self-consistently added to the work of Pancinco et~al.\ (2010) 
and in total analyzed 70 turnoff and SGB stars.   They found C and N abundances from their 
direct measurement of the CH and CN bands and matched them to both the HST photometry of Milone et~al.\ 
(2008) and the Stromgren photometry of L09.  They similarly found that the bright (blue) SGB is 
typically more C-rich and the faint (red) SGB is typically more C-poor, but also that the bright SGB 
is poorer in N and the faint SGB is richer in N.  They argue that there is a significant and possibly 
bimodal spread in the distribution of C+N abundances, with the faint SGB typically being much richer.  

\begin{figure}[htp]
\begin{center}
\includegraphics[scale=0.44]{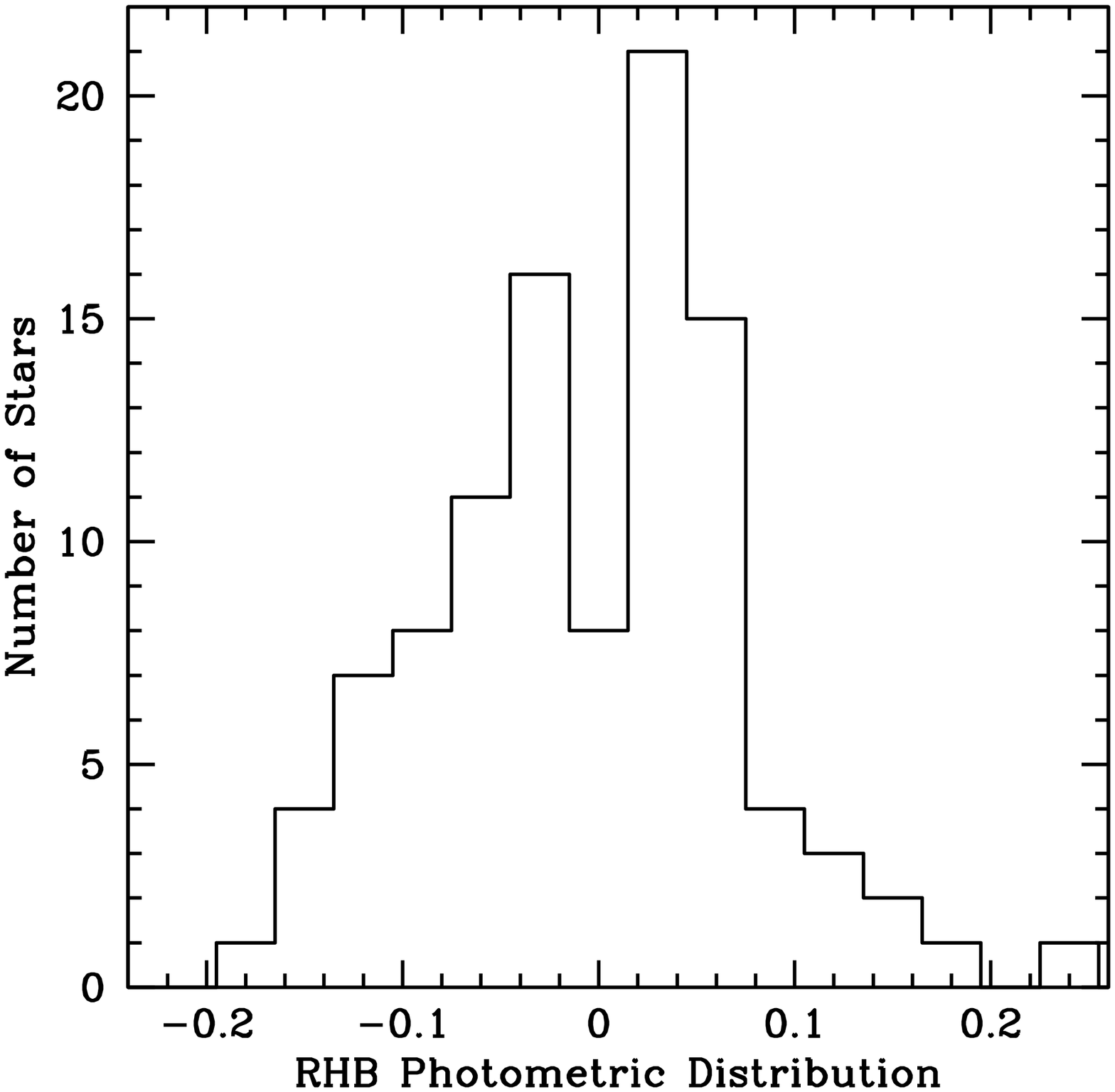}
\vspace{-0.43cm}
\end{center}
\caption{Illustration of the bimodal RHB distribution perpendicular to the solid grey dividing line shown
in Figure 10.}
\end{figure}

G12 question these large variations in the L12 C+N abundances and 
suggest that their T$_{\rm eff}$ were 500 K too cool.  This would cause them to greatly underestimate the C 
abundances, and because the N abundances are found from CN this would also cause the N abundances to be 
overestimated.  However, based on the idea that C+N+O is fairly constant, this variation in C+N would 
at least be qualitatively in agreement with the variations in O abundance observed in the RGB.  The 
C+N-rich stars would 
be O-poor and the C+N-poor stars would be O-rich.  L12 also concluded this by comparing the average RGB 
O-abundances from V10 to their average C+N abundance for the two branches and found no meaningful difference 
between their average C+N+O abundances.  We have also matched 65 of the abundances 
from L12 to our photometry, but only one of these stars could reasonably be defined as belonging to our 
faint SGB.  Most of the limited number of faint-SGB stars from L12 were affected by crowding issues in
our ground-based photometry from Paper I.  The one clear faint-SGB star 
from Paper I with L12 abundances is both N-rich and C-rich, but this is too limited to draw any conclusions.

\section{Matching Horizontal Branch Abundances to Photometry}

The horizontal branch (HB) abundances provide a unique case to analyze the two populations
because here they create two photometrically distinct groups of stars in the blue and red
HB, unlike the overlapping RGBs and SGBs.  It is believed that the BHB corresponds 
to the red and broader population on the RGB and that the RHB corresponds to the blue and narrower
population on the RGB.  The distinct color differences in the HB itself are believed to result from a
moderate He enhancement in the BHB population (e.g., the observations of G12b and the models of Joo 
\& Lee 2013).  This variation in He can also explain the observed variations of the pulsational properties 
of RR Lyrae variables in NGC 1851 (Kunder et~al.\
2013).  Another advantage of the HB is that these are bright stars with relatively small photometric 
error.  Our observations from Paper I were able to observe a broad range of both T1 and T2 
magnitudes in the RHB, and in T1 there is an apparent split that creates two sequences of T1 magnitudes.  
Does this suggest that there may be key differences for stars within the RHB itself?  

Figure 10 shows the abundances from G12b matched to the RHB and BHB (when available).  Similar to previous Figures for
Ca, Fe, and Ba the blue data represent poor stars and the red data represent rich stars for each element.  
However, for the elements that also have BHB abundances measured we used three abundance groups because 
they typically have very broad elemental distributions: for O the red data represent O-rich stars, the 
green data represent moderately O-poor stars, and the blue data represent very O-poor stars.  For Na and 
Mg the blue data represent poor stars, the green data represent moderately rich stars, and the red data 
represent very rich stars (see Figure 10 for the detailed abundance ranges).  

The BHB is very O-poor, Na-rich, and very Mg-rich, consistent with it being the same population that 
creates the red RGB.  
The RHB overall does show a broader range of O, Na, and Mg abundances than the BHB, but there are 
no consistent abundance differences between the two apparent RHB sequences.  While the faint RHB may 
on average be more O-poor, Na-rich, and Mg-rich, it still contains many O-rich, Na-poor, 
and Mg-poor stars.  For both [Fe/H] and [Ca/Fe] there are no meaningful abundance differences between the two 
RHB sequences, but when considering errors the observed distribution spreads in [Fe/H] and 
possibly [Ca/Fe] are not meaningful.  Lastly, Ba appears to similarly show no meaningful difference between 
the bright and faint RHB, but there is a meaningful sample of very 
Ba-rich stars ([Ba/Fe]$>$0.5; see Figure 2) that nearly all fall on the faint RHB.

To look more closely at the double distribution in NGC 1851, we divide the two sequences with the
solid grey line shown in Figure 10 (T1=14.487+(C-T1)$\times$1.156).  In Figure 11 we illustrate the
bimodal distribution perpendicular to this dividing line.  This also provides a reference to statistically
test for differences in photometric distributions on the RHB abundances of Figure 10.  Consistent with 
expectations, there are no significant (p-value $<$ 0.05) differences in distribution of
any of the defined abundance groups within the RHB.  Even the very-rich Na stars in Figure 10 ([Na/Fe]$>$0.23), 
which in the RHB primarily fall on the faint RHB, do not have a statistically meaningful difference
in distribution in comparison to the Na-poorer ([Na/Fe]$\leq$0.23) stars.  This is because while a KS-test provides a moderate 
D of 0.4312, the small number limitations of only 11 very-rich Na RHB stars gives a p-value of
only 0.063.  However, we again make note of the RHB stars richest in Ba ([Ba/Fe]$>$0.5) all primarily
fall on the faint RHB.  They have a meaningfully different photometric distribution in comparison to all
RHB stars with weaker Ba, where a KS-test test provides a D of 0.4701 and a p-value of 0.033.

In comparisons to other cluster, this double sequence in the RHB of NGC 1851 appears similar to the double
sequence in the RHB of 47 Tuc (Milone et~al.\ 2012).  But unlike NGC 1851, 47 Tuc does not also have a BHB.
Another distinction is that these two 47 Tuc RHB sequences are separated in the UV.  Therefore, adopting 
in 47 Tuc two populations with appropriate CNO variations that can create its observed double sequences in 
the MS, SGB, and RGB 
would also create this double sequence in the RHB.  In NGC 1851 its two populations instead create its well 
observed BHB and RHB.  Is this the result of the NGC 1851 populations possibly having a more significant 
difference in He than those of 47 Tuc?  The two RHBs of NGC 1851 are further differentiated
because they are not defined by a difference in UV but by differences in the T1 and T2 magnitudes, 
which are not meaningfully affected by variations in CNO.  

Is there a possible age spread between these
two RHBs?  Milone et~al.\ (2008) suggested an age difference of $\sim$1 Gyr between the two
NGC 1851 populations to explain the observed split SGB, but this corresponds to the RHB and the BHB.
Is there possibly a spread in age between the apparently single population that creates the RHB?
Is there some factor driving a difference in mass-loss rates between the two RHB sequences?
Lastly, is this related to the four, not just two, abundance group spectroscopically observed 
in NGC 1851 by Campbell et~al.\ (2012) and Simpson et~al.\ (2017)?

The apparent double RHB remains a mystery, but the differences between the RHB and BHB have further 
established the key abundance differences between the two populations of NGC 1851.

\section{Photometric Synthesis of Red Giant Branch Populations}

To create our synthetic magnitudes we applied the models of Kurucz et~al.\ (1992) with overshoot 
and used the spectral synthesis program SPECTRUM v2.76\textsuperscript{4} and its linelist (Gray 
\& Corbally 1994).  For our standard input abundances we applied a [Fe/H] of -1.23, which is based 
on that found by V10 (-1.23$\pm$0.01) and is similar to that found by Ca11 ([Fe/H]=-1.16$\pm$0.05).  
We also applied a standard He abundance of Y=0.246.  To analyze the effects of possible 
variations in He, we have performed comparisons to a more He-rich (Y=0.30) population.  For CNO 
abundances we applied the result of V10 that for both populations C+N+O is constant at log 
$\epsilon$(CNO)=8.00.  Additionally, we considered the effects of distinct log $\epsilon$(CNO) in 
the two populations that results from the more significant differences in [N/Fe] measured in Y15.  
This gives the second population a significantly richer log $\epsilon$(CNO)=8.50.  All other 
abundances are scaled solar.  

\begin{figure*}[htp]
\begin{center}
\includegraphics[scale=0.8]{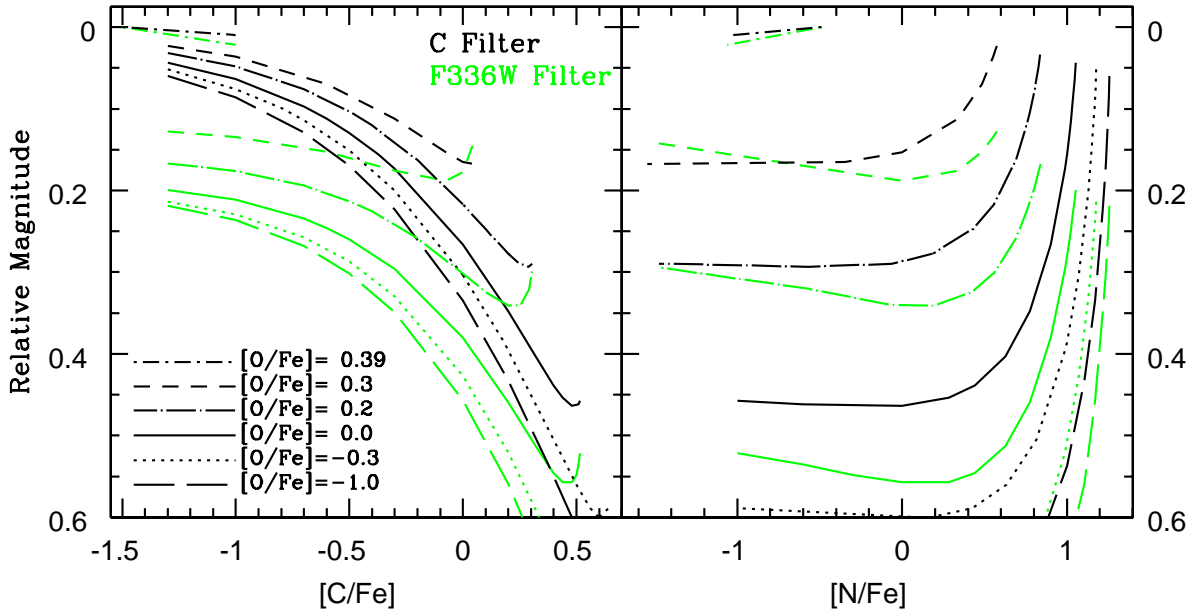}
\vspace{-0.43cm}
\end{center}
\caption{Relative magnitude synthesis of a representative RGB star for the Washington 
C filter (black) and the F336W filter (green).  Varying levels of [C/Fe], [N/Fe], and [O/Fe] are
considered but with an adopted constant log $\epsilon$(CNO)=8.0.  The different curves represent constant
[O/Fe] (see key) with corresponding variations in [C/Fe] (illustrated in the left panel) and [N/Fe] 
(illustrated in the right panel).  Both panels show reflections of the same data curves in [C/Fe] and
[N/Fe] space, respectively.  For both filters their magnitude at the most C-poor abundance synthesized 
is defined as the zero point (brightest) and as standard an increase in magnitude (going down the y-axis) is
increasingly fainter.  This demonstrates the strong dependence of C magnitudes on [C/Fe] across a large range
of abundances.  F336W magnitudes are also strongly dependent on [C/Fe] but are more counterbalanced by an
important dependence on [N/Fe].}
\end{figure*}

Variations in [Fe/H] can also play an important role in MPs, but evidence for a meaningful spread in [Fe/H] in 
NGC 1851 is limited to the analysis in the RGB by Ca11.  V10 did not find any evidence for such a spread.
If there is a true variation it is 
relatively minor ($\sim$0.1 dex).  Additionally, matches of these [Fe/H] values to both Stromgren photometry (see
Ca11) and our own Washington photometry find that these apparent [Fe/H] variations do not match with
the two identified populations and are randomly distributed photometrically.  Therefore, if this minor [Fe/H] 
variation is real it does not play a role in the photometric differences of the two populations, and it would 
only increase the photometric spread observed in each population.  For completeness, however, we briefly 
considered the photometric effects of a 0.1 dex metallicity increase.

\footnotetext[4]{http://www.appstate.edu/~grayro/spectrum/spectrum.html}

\subsection{Effects of CNO on Magnitude at Constant Log $\epsilon$(CNO)}

To thoroughly evaluate the effects of variations of C, N, and 
O, we analyzed nearly the full range of abundances within the constraint of log $\epsilon$(CNO)=8.0.  
This allows us to move beyond the constraints of the abundances of NGC 1851 and also make more general 
conclusions about the photometric effects of CNO variations at constant log $\epsilon$(CNO) and how
this varies in differing filters.

For our representative RGB star we adopted T$_{\rm eff}$=4863 K, log g=2.18, 
microturbulence=1.52 km/s, and [Fe/H]=-1.23.  This places this giant right below the RGB bump in 
NGC 1851.  In addition to the C filter we also synthetically looked at the commonly adopted 
F336W filter (which is very similar to Johnson U).  In contrast to both the C and Stromgren filters, 
Johnson U appears to 
create two more cleanly separated photometric branches in the RGB of NGC 1851 (H09) instead of heavily
overlapping narrow and broad branches.  However, based on published analyses of NGC 1851, it is unclear 
if the U or F336W filters also distinctly separate, for example, the Ba- and Na-poor stars from the Ba- and 
Na-rich stars.  H09 showed that in U-I their red branch is Ca-rich and their blue branch is Ca-poor 
with little meaningful overlap, but these Ca abundances were photometrically based, and Han et~al.\ subsequently
discovered that their Ca filter was old and degraded, leading to an increased sensitivity to CN variations
and a decreased sensitivity to Ca variations.  In other clusters, however, the recent photometric and spectroscopic 
analysis of the similar double RGB in M2 (Lardo et~al.\ 2013) also finds in U-V that the redder RGB is cleanly 
separated from the bluer RGB.  The abundance and spectral indices in Lardo et~al.\ (2013) also illustrate that 
U magnitudes cleanly separate stars of differing Sr and Ba abundances and CH and CN indices.  These 
characteristic differences in how MPs are photometrically distributed in C and U or F336W filters 
was briefly discussed in Paper I, but in this Section and the following we will analyze the reasons for this difference
in more detail.

The general relations between abundance and magnitude for the C and F336W filters can be defined by 
looking at broad abundance ranges for [C/Fe], [N/Fe], and [O/Fe].  Figure 12 assumes a constant 
log $\epsilon$(CNO)=8.00, and the left panel looks at several curves of constant [O/Fe] and how variations in [C/Fe] 
affect both C magnitudes and F336W magnitudes.\footnotetext[5]{In our analysis we adopt for
the Sun a log $\epsilon$(C)=8.49, log $\epsilon$(N)=7.95, and log $\epsilon$(O)=8.83.  This is
consistent with that adopted in V10 and agrees with the Grevesse \& Sauval (1998) solar abundances
within 0.03 dex.}  We remind the reader that variations in [C/Fe] must have corresponding [N/Fe] variations 
in order to keep total CNO constant, which are similarly illustrated in the right panel of Figure 12.  
Throughout this paper we focus on the comparisons plotted relative to [C/Fe] because the 
general characteristics are more clearly displayed in this manner.  The magnitudes are placed on a 
relative scale where the magnitude of the most C-poor and O-rich abundance we analyzed ([N/Fe]=-0.49, 
[C/Fe]=-1.50, and [O/Fe]=0.39) is set as the zero-point for both filters.  Each curve represents a constant 
[O/Fe] with black curves for the C filter and green curves for the F336W filter.  Lastly, we find that 
these same very broad variations in CNO have only marginal effect on the synthetic R (T1) magnitudes
($\lesssim$0.02 mag, or only 2 to 5\%\, of the variations found in C magnitude) but they become more
important for the synthetic I (F814W) magnitudes ($\lesssim$0.04 mag, or 4 to 10\%\, of the variations found in
C magnitude or 2 to 7\%\, of the variations found in F336W).  Therefore, these plotted variations in C and F336W 
magnitude reliably trace variations in C-T1 and F336W-F814W, in particular when looking at realistic 
(moderate) CNO variations observed within a single cluster, but for the highest precision
comparisons to cluster observations these variations in R (T1) and I (F814W) should be noted.

At constant [O/Fe], increasing [C/Fe] with the corresponding decrease in [N/Fe] leads to a 
fainter C magnitude in all but the richest [C/Fe] (weakest [N/Fe]).  Therefore, in the C magnitude 
the [C/Fe] generally plays the dominant role, which was the filter's original purpose (note that the C 
designation in fact refers to Carbon).  [N/Fe] remains important but generally 
plays a secondary role.  In contrast, the F336W magnitude patterns are more complex and typically show at 
constant [O/Fe] a relatively weaker dependence of F336W magnitude on [C/Fe], but followed by quickly brightening
F336W magnitude at the richest [C/Fe].  This pattern can be explained by the F336W filter being strongly dependent 
on [C/Fe] but [N/Fe] plays nearly as important of a role, resulting from the prominent NH band at 3360 \AA\, near
the center of its bandpass and the lack of the CH feature at 4300 \AA\ (the G band).  In contrast the G band 
plays a more important role in the C filter while the strong NH band is at the far blue edge of its bandpass
where the transmission is relatively low.  
Therefore, for the F336W filter in Figure 12 the increasing [C/Fe] at constant [O/Fe] is counterbalanced by a 
decreasing [N/Fe], leading to a comparatively weaker magnitude change with increasing [C/Fe] followed by 
a significant brightening at the C-rich extreme because so little N remains, greatly weakening the 
NH line.

\begin{figure*}[htp]
\begin{center}
\includegraphics[scale=0.85]{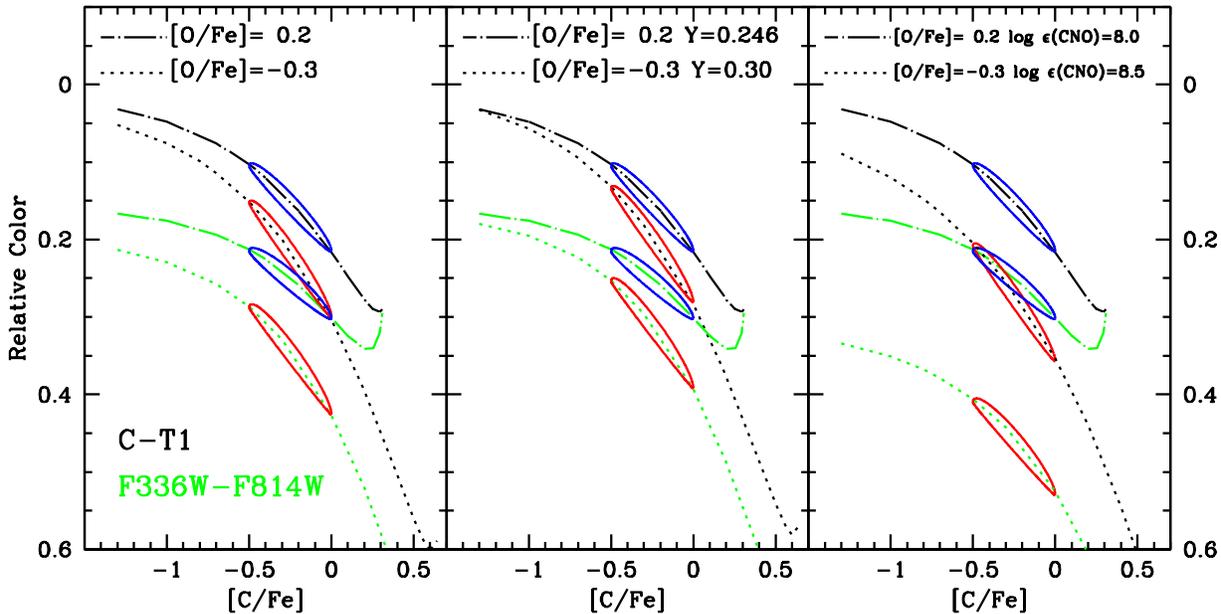}
\vspace{-0.43cm}
\end{center}
\caption{The left panel applies the curves from Figure 12 to C-T1 and F336W-F814W color space for 
abundances of the two populations in NGC 1851.  An O-rich/N-normal population is circled in blue and 
an O-poor/N-rich population is circled in red for both color curves.  This illustrates
photometric variations similar to the corresponding observations of Paper I for C-T1 (a narrow blue 
population and a heavily overlapping but broad red population) and H09 for U-I (which is comparable to
F336W-F814W with two more cleanly separated populations).  The central panel looks at identical CNO variations 
but illustrates the effects of
O-poor/N-rich population ([O/Fe]=-0.3) also being He-rich (Y=0.30).  This causes this redder population to
be shifted slightly bluer, but the two populations are still distinct.  The right panel looks at the significant
effect of the O-poor/N-rich population ([O/Fe]=-0.3) also having a richer log $\epsilon$(CNO)=8.5.  The significant 
increase in [N/Fe] shifts the redder populations $\sim$0.06 further to the red in C-T1 and $\sim$0.12 further
to the red in F336W-F814W.}
\end{figure*}

Comparing the relative shifts in the differing lines (representing changing [O/Fe]) indicates 
another key difference between the two filters caused by the stronger [N/Fe] dependence of F336W.  
With increasing [O/Fe], at constant [C/Fe], the F336W magnitudes become fainter more rapidly 
than the corresponding C magnitudes.  This is not directly the result of the changing [O/Fe], which
plays relatively little importance in the actual flux, but the corresponding increase in [N/Fe] with 
the decreasing [O/Fe].  This also represents the advantage that the F336W (and the similar Johnson U) 
filter has over the C filter in separating MPs that have a significant difference in [N/Fe].  It
is also important to note an anti-correlation (or correlation) between NH and CH strengths is not clearly
observed in NGC 1851, but they are observed in most globular clusters (e.g., Pancino et~al.\ 2010; L15).  For
clusters with anti-correlations (e.g., 47 Tuc, M15, NGC 288) the weaker CH bands in the N-rich/O-poor 
second population will weaken the observed magnitude shift in C more than it does in F336W, but for clusters 
with a positive correlation (e.g., M22) this will increase the C magnitude sensitivity more than for F336W.
Despite these caveats, the C filter 
is still highly sensitive to CNO abundance variations and retains many advantages with its significant 
(4 to 5 times) increase in photometric throughput in cool RGB stars versus the F336W filter.  

\subsection{Application to NGC 1851 Red Giant Branch Abundances}

Here we apply these curves more specifically to the photometric and spectroscopic observations of NGC 1851.
In Figure 13, because we look at a number of special cases where variations in R (T1) and I
(F814W) play a role, we plot these curves in color space with C-T1 in black and F336W-F814W in green. 
In the left panel of Figure 13 we have selected two appropriate abundance ranges for the two observed 
populations of NGC 1851.  Both populations have a comparable average [C/Fe] of $\sim$-0.25 but with a 
moderate variation, consistent with no CN-CH correlation.  The dominant blue
population has a [O/Fe]$\sim$0.2 while the secondary red population has a [O/Fe]$\sim$-0.3.  This is more 
C-rich than the RGB abundances measured by V10 and Y15, but those are from upper RGB stars
where deep mixing has further depleted the [C/Fe] beyond the lower RGB and the SGB.  For
both the C-T1 and F336W-F814W curves we illustrate the abundance groups with a blue and red ellipse plotted over
each curve set.  As would be expected, the O-rich/N-normal population (blue ellipse) has both brighter C and F336W 
magnitudes giving relatively narrow and blue populations in both colors.  Also, as expected, the O-poor/N-rich 
population (red ellipse) has both fainter C and F336W magnitudes, giving redder colors, but there are many key 
differences.  First, in C-T1 the weaker [N/Fe] sensitivity leads to a relatively weaker color 
difference causing strong overlap between the primary and secondary population at the O-poor/C-poor edge versus
the O-rich/C-rich edge.  However, the sensitivity of C-T1 to varying [C/Fe] is increased in the O-poor 
stars giving a very broadly distributed color range with a distinctly redder subpopulation consisting of only 
the O-poor/N-rich/C-rich stars.  The second population (O-poor/N-rich) in F336W-F814W shows more significant 
color separation from the primary blue population.  Additionally, the weaker sensitivity to [C/Fe] variations 
leads to a narrower population than observed in C-T1.  These findings are strongly consistent with our
observations in Paper I of the two RGB branches in C-T1 and C-T2 and the comparable U-I used by H09. 

\begin{figure*}[htp]
\begin{center}
\includegraphics[scale=0.85]{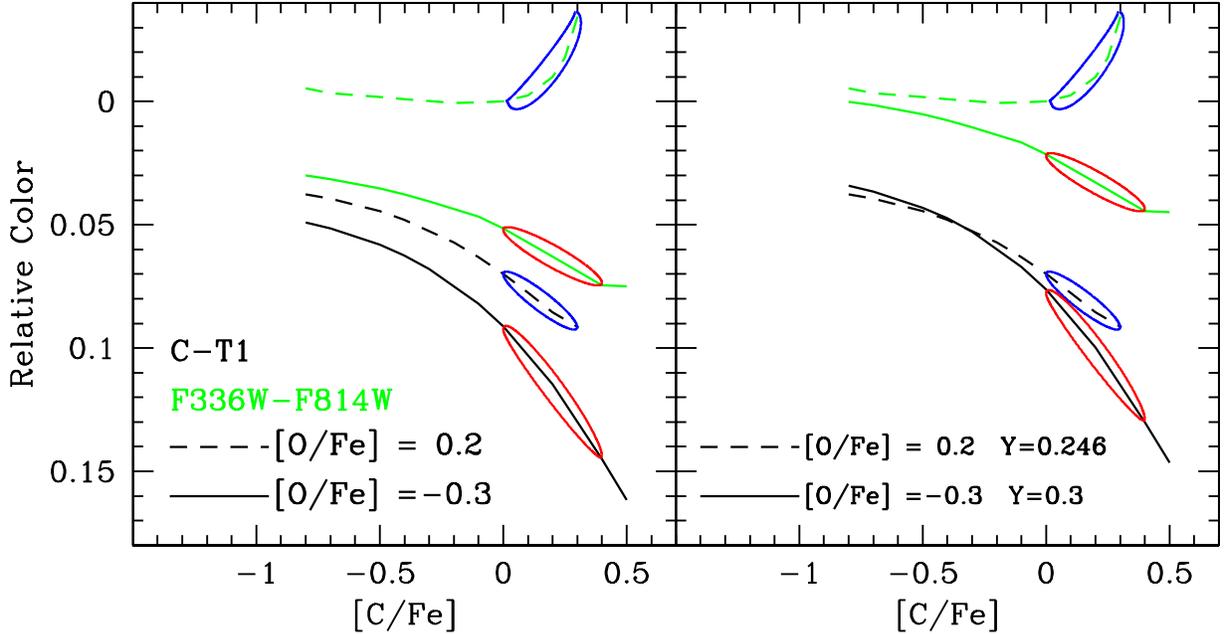}
\vspace{-0.43cm}
\end{center}
\caption{The left panel shows the relative color synthesis of a representative upper MS star for 
C-T1 (black) and for F336W-F814W filter (green).  The format is the same as that described in Figure 13.  
Additionally, the C-T1 and F336W-F814W curve sets are arbitrarily offset for clarity.  While the [C/Fe] 
sensitivity of these colors are weaker here than in the RGB, due to the hotter T$_{\rm eff}$, it is still 
strong and the trends are similar to that found in the RGB.  We note that in the MS these same two populations 
should overall be more C-rich and correspondingly poorer in N than the RGB.  Two photometrically separated 
populations are predicted in both colors.  A double MS, however, has not been observed in F336W-F814W for NGC 
1851.  In the right panel we consider that the redder (O-poor/N-rich) population is likely also 
He-rich.  Adjusting the relative colors for an He difference
greatly diminishes the predicted photometric difference in F336W-F814W, and with only moderate photometric 
error these populations may appear as a somewhat broad but single MS.  The C-T1 color difference
is also diminished, but the O-poor/N-rich population remains significantly broader in color and can be detected
more easily through the distribution analysis performed in Paper I.}
\end{figure*}

The consistency of the general photometric characteristics seen in observations is promising.
More quantitatively, the observations in both Paper I and H09 find an observed difference of 
$\sim$0.1 in C-T1 and U-I colors, 
respectively, between the centers of the two RGB branches.  This is consistent with 
the separation of the centers of the red and blue ellipses in F336W-F814W color.  For C, while the
separation of the two ellipses in C-T1 color is $\sim$0.05, this is the separation of the two
complete and overlapping populations.  The RGB ``red branch" in C-T1 is only the redder half
of the red ellipse, and again it has an extension of $\sim$0.1 beyond the dominant blue branch.
(For reference, the O-poor/N-rich population is only $\sim$0.005 magnitudes fainter in R (T1) 
and $\sim$0.015 magnitudes fainter in I (F814W; T2) in comparison to the O-rich/N-poor population.  
The lack of significant effect of CNO variations on these redder filters is also consistent with
the two RGB branches not being observed in Paper I's T1-T2.)  

While our photometric matches do not find that redder RGB population is more metal-rich, it is important
to test what effects metallicity variations may have between the two populations.  Adopting that the
O-poor/N-rich population is also 0.1 dex richer in all metals (besides CNO) finds that all filters are moderately
affected.  Accounting for the effects in both filters, the metallicity increase would shift the O-poor/N-rich 
population $\sim$0.02 redder in C-T1 and $\sim$0.025 redder in F336W-F814W.  By itself, this metallicity increase
is not large enough to create the observed color differences between the two populations in NGC 1851, but
this shows that metallicity is important to consider in clusters with well defined metallicity differences 
between their MPs.  As suggested by the NGC 1851 observations in Ca11, however, if there is a true metallicity 
spread of $\sim$0.1 dex within each population, this would further enhance the color spread within each population.

In the center panel of Figure 13 we also consider the effects of the second population (O-poor/N-rich)
being more He-rich (Y=0.3) than the first population.  In an RGB star of these characteristics, this
has minor but measurable effects on its F336W and C magnitudes.  However, we note that variations in He 
also have smaller but meaningful effects on the R (T1) and 
F814W (I) magnitudes.  Therefore, in the center panel of Figure 13 we account for all filter variations
in our color curves.  As expected, a richer He shifts the both C-T1 and F336W-F814W to bluer colors.

In the right panel of Figure 13 we look at the effects that an increase of log $\epsilon$(CNO) from 8.0 
to 8.5 for the second population would have.  This is driven by a significant increase in [N/Fe] but with 
consistent [O/Fe] and [C/Fe] abundances and is in line with with the abundances from Y15.  Unlike 
at constant log $\epsilon$(CNO), this variation in total log $\epsilon$(CNO) leads to more important effects 
on R (T1) at $\sim$0.03 and F814W (I) at $\sim$0.05.  Therefore, we apply the effects of all filters on our 
colors.  We find that this large increase in [N/Fe] causes a significant shift 
to the red for the second population that is far more than is found in either C-T1 or U-I observations.  Even when 
correcting for the second population also being richer in He, this remains inconsistent 
with observations.  One additional factor suggested by the abundances of V10 and G12 is that the second
population on average is slightly ($\sim$0.1 dex) more C-poor.  This would help mitigate the effect of 
such a large increase in [N/Fe] and suggests that the log $\epsilon$(CNO) may not be uniform across both 
populations, but any differences are likely more moderate ($\lesssim$0.2-0.3 dex).

\section{Photometric Synthesis of Main Sequence Populations}

In Paper I, using Washington photometry we discovered evidence for NGC 1851 having two photometric branches 
in the MS consistent with those observed in its RGB.  Unlike in the other stellar
groups, there are no direct spectroscopic abundances available for MS stars in NGC 1851.
However, we can still infer general abundance information from the observed SGB and RGB abundances.  
Therefore, we have looked specifically at an O-rich ([O/Fe]=+0.2) population 
and an O-poor ([O/Fe]=-0.3) population, consistent with our RGB analysis and the abundances
of Ca11.  At these [O/Fe] we again adopt a constant log $\epsilon$(CNO)=8.0 and at a range of 
[C/Fe] near scaled solar.  Our representative MS model has T$_{\rm eff}$=5800 K, log g=4.59, 
microturbulence=0.74 km/s, and [Fe/H]=-1.23.  This places our star in the upper MS below the turnoff.

Figure 14 looks at the color synthesis for an MS star in both the C and F336W filters.  Both the general 
C-T1 and F336W-F814W color trends are consistent with that found in the RGB (Section 6).  Again, 
these color variations are driven by magnitude variations in C and F336W because the synthetic R (T1) and 
F814W (I) magnitudes variations ($<$0.01) are even weaker than those predicted in the RGB.  Overall, in
comparison to the RGB, the variations in synthetic color are weaker but still significant.  This is primarily 
the result of the MS model star being $\sim$1000 K hotter than the RGB model star, giving weaker CN, CH, 
and NH bands, rather than any other difference between MS and RGB stars.  

Focusing on the expected abundance differences in the two MS populations of NGC 1851,
the [C/Fe] abundances of the SGB and upper turnoff stars from G12 (see Figure 1) suggest that the [C/Fe] in the MS
will have moderate variations and be richer than scaled solar.  The MS is expected
to be more C-rich than the RGB because the mixing processes that dilute the surface C during evolution
have not yet begun.  In Figure 14 we have adopted
abundance ranges consistent with this for the two MS populations and find that the O-rich/N-normal
population (blue ellipse) is bluer in both C-T1 and F336W-F814W and relatively narrow, but  
at these very C-rich abundances the decreasing [N/Fe] makes the F336W-F814W colors increasingly bluer.
The O-poor/N-rich population (red ellipses) create in C-T1 a very broadly distributed
and slightly redder population with minor overlap with the blue population.  The F336W-F814W creates a 
significantly redder but still photometrically narrow population.

Milone et~al.\ (2008) did not find evidence in NGC 1851 for a 
double MS in F336W-F814W, but the C-T1 observations from Paper I do find evidence for a heavily overlapping 
but broadly distributed second redder MS.  In the upper MS this second MS extends to the red with an apparent separation 
of $\sim$0.05 magnitudes.  In the lower MS this separation increases further, resulting from the increasing
strengths of the molecular bands in cooler stars.

Figure 14 suggests that the double MS should be cleanly separated in the F336W-F814W observations of Milone 
et~al.\ (2008), but this is not the case.  A possible explanation for this is that in the MS the effect of the 
varying He abundances in the two populations becomes more important.  The O-poor/N-rich population
is believed to be more He-rich.  In the right panel of Figure 14 we analyze the effects of this by keeping 
the O-rich/N-normal population at our standard Y=0.246 and the O-poor/N-rich population
at Y=0.30.  Across the range of [C/Fe] and [N/Fe] at constant [O/Fe]=-0.3, the effect on the C and F336W
magnitudes are uniform at 0.043 brighter and 0.052 brighter, respectively.  The He-rich population is also 
0.028 magnitudes brighter in T1 and 0.022 magnitudes brighter in F814W.  This shifts the He-rich population 
0.03 magnitudes bluer in F336W-F814W and 0.015 magnitudes bluer in C-T1.  This brings the populations in
C-T1 closer together but because of their distinct color distributions the heavily blended redder 
population was still observed in Paper I using C-T1.  However, the larger effect of He on the F336W-F814W 
color nearly brings the two narrow MS populations together.  With even small photometric error and
considering that this second population may be slightly more C-poor ($\sim$0.1 dex), this may appear 
similar to a single MS that is broad in color but otherwise normal.

\section{Summary \& Conclusions}

Matching previously published spectroscopic abundances to our Washington photometry, we have analyzed many of the quantitative 
differences between the two NGC 1851 populations.  The large sample of RGB abundances from Ca11 have shown 
that Ba, Na, and O are useful elements to distinguish between the two populations in the RGB.  In particular, 
Ba and Na clearly show that stars poor in either of these elements are consistently part of the (in C-T1) narrow 
and blue RGB, while stars that are rich in either of these elements are more broadly distributed in color.
The reddest stars of this second population create the red RGB while the bluer stars of this second population
are consistent in color with the blue RGB.  In agreement with the strong O and Na anti-correlation, the 
O-rich stars are consistent with the narrow and blue RGB while the O-poor stars are more broadly distributed.  
To better understand this broad color distribution of the second population, we considered two abundances 
simultaneously.  The stars that are both Ba-rich and O-poor (or similarly Ba-rich and Na-rich) 
all fall primarily on the red RGB.  All other stars are consistent with the 
blue RGB.  This may result from [Ba/Fe] acting as a tracer of [N/Fe].  Therefore, the red RGB is composed of 
only the N-rich/O-poor stars.  This is further supported because stars that are both CN-rich
and Na-rich (or CH-rich) all fall primarily on the red RGB.

Matching the G12 abundances to the two branches of the SGB shows many consistencies with what we found in 
the RGB abundances, in particular for the correlated elements of Ba and Sr.  While the lack of O and Na 
abundances for the SGB limit our analysis tools, the C abundances from G12 are useful to analyze these two 
branches.  Most interestingly, the C abundances from G12 suggest that the faint SGB is more C-poor and the 
bright SGB is more C-rich, but the difference is only $\sim$0.1 dex while the total range of C abundances 
span $\sim$0.6 dex.

For the G12b abundances of the HB, as expected we find clear abundance differences between the RHB and BHB 
where the BHB is relatively very O-poor and more Na-rich and Mg-rich.  However, other than a group
of very Ba-rich stars being predominantly in the faint RHB, there are no clear abundance 
differences between the observed bright and faint RHB sequences.  The cause of these sequences in the T1 and 
T2 magnitudes remains unclear, but it is likely different than the CNO and He variations that can explain the 
two established populations, represented by the BHB and the RHB.  It would be of interest to investigate what 
other factors may cause two sequences of such characteristics in the ``single population" of the RHB.

To build on these abundance analyses and look at the role that CNO plays in the C and F336W filters, we performed 
photometric synthesis for a broad range of [C/Fe], [N/Fe], and [O/Fe] at a constant log $\epsilon$(CNO)=8.0.
Comparisons of the effects on both the C and F336W filters find that at constant C+N+O the variations in
magnitude can be quite significant and adopting larger log $\epsilon$(CNO) values are not necessary
to explain the observed photometric differences between the two populations.  This analysis also showed that
the C filter is highly sensitive to [C/Fe], as its name would suggest, with some sensitivity to [N/Fe].
In contrast, F336W and the similar Johnson U filter have more comparable sensitivities to both [C/Fe] 
and [N/Fe].  

To compare these synthetic magnitudes to observations of NGC 1851, we adopted abundance characteristics for 
each population based on the RGB spectral analyses from Ca11, V10,
Ca14, and Y15.  We define the two populations as an O-rich ([O/Fe]=0.2) and an O-poor ([O/Fe]=-0.3) population.
In our representative RGB star that is just below the RGB bump, we assume that both populations have comparable 
but broadly distributed [C/Fe] near [C/Fe]$\sim$-0.25, which is consistent with NGC 1851 not having
a CN-CH correlation.  Based on the constant C+N+O, and consistent with the CN observations of
Ca14, the O-rich and O-poor populations are also N-normal and N-rich, respectively.  Application of this
to our photometric synthesis finds that the O-rich/N-normal population is blue and narrowly distributed 
in color in both C-T1 and F336W-F814W.  Similarly, the O-poor/N-rich population are both redder in
C-T1 and F336W-F814W.  However, in C-T1 the second population's colors are very broadly distributed 
and overlapping with the bluer population.  In F336W-F814W the second population is redder and nearly 
as narrow in color as the blue population.  These color variations are driven almost completely by variations
in the C and F336W filters and are consistent with the photometric observations of both Paper I and H09.  

In application to general globular cluster observations, it is important
to discuss the differences between clusters with a CN-CH anti-correlation or correlation in contrast to our
adoption of no correlation (as observed in NGC 1851).
In general, the photometric sensitivity to MPs will be decreased in clusters with CN-CH 
anti-correlations, but more so in the C filter versus F336W.  However, for clusters with CN-CH correlations the 
increase in photometric sensitivity to MPs will be greater in the C filter versus F336W.  Our broad abundance
analysis in Figure 12 helps illustrate this.

We also looked at additional differences that may play a role in the photometric differences of these two 
populations.  First, we considered the O-poor/N-rich population being He-rich (Y=0.3).  This diminishes the 
color difference between the two populations, but the effect is relatively minor and the predicted total 
photometric differences are still significant and detectable.  Second, we considered the 
O-poor/N-rich population being significantly more N-rich than the primary population at log $\epsilon$(CNO)=8.5.  
This greatly increases the predicted color differences between the two populations, strikingly
in F336W-F814W, and well beyond the differences that have been observed in NGC 1851.  A more moderate difference 
in log $\epsilon$(CNO) ($\lesssim$0.2 to 0.3) for the two populations, however, could be further mitigated
by the O-poor/N-rich population being both He-rich and slightly more C-poor ($\sim$0.1 dex).  This could bring 
two populations of a meaningfully different log $\epsilon$(CNO) into agreement with the photometric observations.

Applying the same [O/Fe] based populations to photometric synthesis in the MS finds further interesting
comparisons between the C and F336W filters.  Because the MS will be more C-rich we adopted a moderately spread
[C/Fe] consistent with that found in the least unevolved stars from G12 ([C/Fe]$\sim$0.2).  
In the hotter MS stars, compared to the cooler RGB stars, the weaker molecular bands lead to comparable 
dependencies but overall less significant sensitivity to the CNO abundances in both the C and F336W filters.  
At these richer [C/Fe] the second, redder population is still very broadly distributed in the C-T1 but quite narrow  
in F336W-F814W.  This is consistent with the observations of a heavily overlapping double MS from
Paper I, but how does it explain the lack of observed double MS in the F336W-F814W observations of Milone et~al. (2008)?
We believe this again relates to the increase in He abundance in the second O-poor/N-rich population.
In the MS the increase in He has a more important effect than in the RGB, and it is also stronger in 
the F336W-F814W colors than the C-T1 colors.  This He correction is significant enough that in F336W-F814W 
the two narrowly distributed in color MS populations are nearly shifted on top of each other.  Considering appropriate 
photometric error and the second population being slightly more C-poor ($\sim$0.1 dex), this could create 
a broad but apparently single MS.  In C-T1, the significant difference between the color distribution widths of 
the two populations leaves a comparably overlapping but more distinguishable second population. 

The C filter is a very important filter in analyzing the MPs in GCs for four reasons.  The first is that in 
the typically very cool stars in GCs the C filter has $\sim$3 to 5 times the throughput of other UV filters.  
This results from the C filter being both broader but also centered redward of the other UV filters, which 
makes it less susceptible to reddening and interstellar extinction.  Lastly, the C-filter's peak sensitivity 
is also usually higher than its competitors.  These factors combined allow MPs to be relatively quickly 
identified using smaller telescopes, like in the analysis from Paper I of NGC 1851 using relatively little 
time on a 1-meter telescope.  This increased sensitivity also greatly improves our ability to observe distant 
populations in both our own Galaxy and also in the Magellanic Clouds, as well as the Andromeda Galaxy with 
the HST.  Second, the C filter may provide much stronger sensitivity than the F336W filter for detecting 
multiple MSs, at least for populations like that found in NGC 1851, but potentially for many other GCs as 
well.  Third, while both the C and F336W filters can detect MPs, we have shown several important differences 
in how these two magnitudes are affected by the compositional variations expected in MPs.  Analyzing GCs with 
both filters and comparing how their MPs are distributed differently in both filters provides an efficient
method to help constrain the abundance differences between the two populations, in particular when 
spectroscopic abundances may be limited.  Finally, although the combined ``magic trio" of F275W, 
F336W, and F438W filters on board HST are unmatched in their ability to distinguish MPs, we will be forced to 
use a different technique once HST has stopped operations.  We will be ''uv-blind" with no ability to observe 
blueward of the atmospheric cutoff and ground-based filters like Washington C will be especially important.

Acknowledgements: J.C. acknowledges Christian Johnson for his help with spectral synthesis software
and support from the National Science Foundation (NSF) through grant AST-1211719. 
D.G. and S.V. gratefully acknowledge support from the Chilean 
BASAL Centro de Excelencia en Astrof\'isica y Tecnolog\'ias Afines (CATA) grant PFB-06/2007.

\end{document}